\documentclass[nofootinbib,preprint,amsmath,amssymb,aps,eqsecnum]{revtex4}
\usepackage[utf8,latin1]{inputenc}
\usepackage{graphicx}
\usepackage{dcolumn}

\usepackage{mathrsfs}  
\usepackage{cases}
\usepackage{bm}
\newcommand{\qm}[1]{``#1''}
\usepackage{hyperref}
\hypersetup{colorlinks, linkcolor={red},citecolor={blue},urlcolor={blue}}  

\newcommand{\dd}{{\rm d}}

\def\nn{\nonumber}

\begin{document}

\title[Geodesic motion in Euclidean Schwarzschild geometry]{Geodesic motion in Euclidean Schwarzschild geometry}

\author{Emmanuele Battista$^{1}$\vspace{0.5cm}}\email{emmanuele.battista@univie.ac.at}\email{emmanuelebattista@gmail.com}
\author{Giampiero Esposito$^{2,3}$}
\email{gesposit@na.infn.it}

\affiliation{$^1$ Department of Physics, University of Vienna, Boltzmanngasse 5, A-1090 Vienna, Austria \\ 
$^2$ Universit\`a degli Studi di Napoli ``Federico II'',  Dipartimento di Fisica ``Ettore Pancini'', Complesso Universitario 
di Monte S. Angelo, Via Cintia Edificio 6, 80126 Napoli, Italy\\
$^3$ Istituto Nazionale di Fisica Nucleare, Sezione di Napoli, Complesso Universitario 
di Monte S. Angelo, Via Cintia Edificio 6, 80126 Napoli, Italy
}

\date{\today} 

\begin{abstract}
This paper performs a systematic investigation of geodesic motion in Euclidean Schwarzschild 
geometry, which is studied in the equatorial plane. The explicit form of geodesic motion
is obtained in terms of incomplete elliptic integrals of first, second and third kind. 
No elliptic-like orbits exist
in Euclidean Schwarzschild geometry, unlike the corresponding Lorentzian pattern. 
Among unbounded orbits, only unbounded first-kind orbits are allowed, unlike general relativity
where unbounded second-kind orbits are always allowed. 
\end{abstract}

\maketitle

\section{Introduction}

Ever since Schwarzschild obtained his spherically symmetric solution of vacuum
Einstein equations \cite{S1916}, the resulting spacetime geometry has been investigated 
with a huge variety of perspectives. In particular, we find it important to
mention the following works.
\vskip 0.3cm
\noindent
(i) The Regge-Wheeler proof \cite{RW1957} that a Schwarzschild singularity will
undergo small vibrations about the spherical form and will therefore remain
stable if subjected to a small nonspherical perturbation.
\vskip 0.3cm
\noindent
(ii) The detailed investigation of geodesic motion in the case of Lorentzian
signature of the metric performed in Refs. 
\cite{Hagihara1931,Darwin1959,Darwin1961,Chandrasekhar1983}, as well as  the more recent works regarding Schwarzschild-(anti-)de Sitter spacetimes, BTZ black holes, noncommutative Schwarzschild black holes, and static and spherically symmetric traversable wormholes geometries \cite{Hackmann2008,Giri2021,Giri2022a,Giri2022b}. 
\vskip 0.3cm
\noindent
(iii) The proof in Ref. \cite{D2021} that general vacuum initial data with
no symmetry assumed, if sufficiently close to Schwarzschild data, evolve to
a vacuum spacetime which possesses a complete future null infinity, remains
close to Schwarzschild in its exterior, and approaches a member of the
Schwarzschild family as an appropriate notion of time goes to infinity.
\vskip 0.3cm
\noindent
(iv) The work on gravitational instantons in Euclidean quantum gravity
\cite{GH1979}, until the recent discovery of a new asymptotically flat instanton
\cite{CT2011}, and the even more recent proof that all known gravitational
instantons are Hermitian \cite{AA2021}.
\vskip 0.3cm
\noindent
(v) A broader set of investigations in Euclidean Schwarzschild, including 
zero modes \cite{E1}, black holes in matrix theory \cite{E2}, Yang-Mills
solutions \cite{E3,E4}, the master equations of a static perturbation \cite{E5},
multiplicative noise \cite{E6}.
\vskip 0.3cm
\noindent
(vi) The work by the authors in Ref. \cite{BE2021}, where a basic integral
formula of geometric measure theory has been evaluated explicitly in the
relevant case of Euclidean Schwarzschild geometry, and it has been suggested
that the in-out amplitude for Euclidean quantum gravity should be evaluated 
over finite-perimeter Riemannian geometries that match the assigned data 
on their reduced boundary. This work has also obtained a heuristic derivation
of a formula expressing a correction to the classical entropy of a
Schwarzschild black hole. Furthermore,  in Ref. \cite{BE2022} we have provided explicit examples for the concept of generalized discontinuous normals to  finite-perimeter sets in non-Euclidean spaces and two-dimensional gravity settings. 

Motivated by our original calculations in Refs. \cite{BE2021,BE2022}, in this paper we study  
geodesic motion in Euclidean Schwarzschild geometry. The present work can be seen as a  
step towards a novel perspective on some features of classical and quantum Euclidean gravity.
From the point of view of functional-integral quantization, the work in Refs. \cite{BE2021,BE2022}
has in our opinion good potentialities because 
measurable sets belong to two broad families: either
they have finite perimeter, or they do not. In the former case, the tools of
geometric measure theory \cite{Maggi} might help in putting on firm ground the so far purely 
formal work of theoretical physics literature.

If one tries to understand the very nature of quantum field theory, one may
still regard it as integration over suitable function spaces
\cite{Glimm}, at least in order to define and evaluate in-out amplitudes.
This involves the action functional and the effective action, and is 
therefore a part of the relativistically invariant, space-time approach
to quantum field theory \cite{DW1984,DW2003}. The Euclidean approach
is a mathematical framework where this form of quantization acquires a
mathematical meaning and is therefore physically relevant, despite the
fact that the space-time metric has Lorentzian (rather than Riemannian)
signature. For example, one first solves a heat equation for a suitable
Green function, and its analytic continuation yields eventually
the Feynman propagator.

Gravitational instantons play a role in the tree-level evaluation of
quantum amplitudes, and their investigation in the seventies led also
to new results in Riemannian geometry \cite{Pope1981}.

In recent years, some authors have considered a novel geometric 
perspective on the nature of particles. When compact gravitational instantons
are studied, it turns out that the neutron can be described by complex
projective space $CP^{2}$ \cite{Atiyah2012} with the associated
Fubini-Study metric, but more recently \cite{S2015, S2016}, asymptotically
flat instantons such as Euclidean Schwarzschild have been considered as
candidates for a geometric description of the neutron. Although none of these
arguments is compelling, they add evidence in favour of gravitational 
instantons having good potentialities, if one is interested in 
foundational and qualitative features of the laws of nature.

Moreover, the systematic proof of geodesic completeness of gravitational instantons 
as a possible criterion for their singularity-free nature has not been 
attempted nor obtained in the literature, as far as we know. This would 
be of interest both in mathematical and in theoretical physics of fundamental
interactions.

The paper is organized as follows. Section \ref{Sec:Geodesics-Euclidean-Schwarzschild-geometry} obtains the equations for geodesic motion in the equatorial plane. Section \ref{Sec:roots-of-cubic} solves the cubic equation for turning points and provides a qualitative analysis of the orbits, whereas the explicit solution in terms of elliptic integrals jointly  with its graphical representation is obtained in Sec. \ref{Sec:Sol-elliptic.intregrals}. The lack of circular orbits is proved in Sec. \ref{Sec:lack-circular-orbits}. Concluding remarks are made in Sec. \ref{Sec:Conclusion}, and relevant details are given in the appendices.

\section{Geodesic equations in Euclidean Schwarzschild geometry}
\label{Sec:Geodesics-Euclidean-Schwarzschild-geometry}  

The Euclidean Schwarzschild metric expressed in Schwarzschild coordinates 
$(\tau,r,\theta,\phi)$  reads as 
\cite{GH1977,G1983,Esposito1994}
\begin{equation} \label{Schwarzschild-metric-1}
g_E^{(1)}= \left(1-\frac{2M}{r}\right) {\rm d} \tau \otimes {\rm d} \tau 
+ \dfrac{{\rm d} r \otimes {\rm d} r}{\left(1-\frac{2M}{r}\right)} 
+ r^2 \left( {\rm d} \theta \otimes {\rm d} \theta + \sin^2 \theta \; 
{\rm d} \phi \otimes {\rm d} \phi \right),
\end{equation}
where the link with the Lorentzian-signature metric is obtained 
by setting $\tau = {\rm i}t$.
We work on the real Riemannian section where the metric is positive-definite.
This implies that the $r$ coordinate must obey the restriction 
\begin{equation}
r \geq 2M,
\label{r-bigger-2M}
\end{equation}
which agrees with the restriction obtained on using Kruskal-Szekeres coordinates.
Thus, the Kretschmann invariant $R^{\mu \nu \sigma \rho} R_{\mu \nu \sigma \rho}$ is a bounded function on the real Riemannian 
section of Euclidean Schwarzschild.

By exploiting the symmetries of Schwarzschild geometry, we can limit our investigation 
to the equatorial plane $\theta = \pi/2$, where the geodesic equations read as
\begin{subequations}
\begin{align}
& \dfrac{{\rm d}^2 r  }{{\rm d} \lambda^2 } - \dfrac{A^\prime}{2A} 
\left(\dfrac{{\rm d} r }{{\rm d} \lambda }\right)^2 -rA 
\left(\dfrac{{\rm d} \phi }{{\rm d} \lambda }\right)^2 -{AA' \over 2}
\left(\dfrac{{\rm d} \tau }{{\rm d} \lambda }\right)^2=0,
\label{Schwarzschild-geodetica-r}
\\
& \dfrac{{\rm d}^2 \phi  }{{\rm d} \lambda^2 } +\dfrac{2}{r} \dfrac{{\rm d} r }{{\rm d} \lambda } 
\dfrac{{\rm d} \phi }{{\rm d} \lambda } =0, 
\label{Schwarzschild-geodetica-phi}
\\
& \dfrac{{\rm d}^2 \tau  }{{\rm d} \lambda^2 } +\dfrac{A^\prime}{A} \dfrac{{\rm d} r }{{\rm d} \lambda } 
\dfrac{{\rm d} \tau }{{\rm d} \lambda } =0, 
\label{Schwarzschild-geodetica-tau}
\end{align}
\end{subequations}
where $\lambda$ is the affine parameter, the prime denotes the derivative with respect to the 
$r$ variable and we have set
\begin{equation}
A(r) \equiv 1-{2M \over r}.
\end{equation}

After dividing Eqs. (\ref{Schwarzschild-geodetica-phi}) 
and (\ref{Schwarzschild-geodetica-tau}) by 
${\rm d }\phi/{\rm d }\lambda$ and ${\rm d }\tau/{\rm d }\lambda$, respectively, we obtain
\begin{align}
\dfrac{{\rm d}}{{\rm d}\lambda} \left[ \log \left(\dfrac{{\rm d}\phi}{{\rm d}\lambda} 
\right) + \log r^2 \right] &=0,
\\
\dfrac{{\rm d}}{{\rm d}\lambda} \left[ \log \left(\dfrac{{\rm d}\tau}{{\rm d}\lambda} 
\right) + \log A \right] &=0,
\end{align}
from which we derive
\begin{align} \label{constant_1}
\dfrac{{\rm d} \tau}{{\rm d}\lambda} &=\dfrac{C}{A(r(\lambda))}
\\
r^2 \dfrac{{\rm d} \phi}{{\rm d}\lambda} &=J, 
\label{constant_2}
\end{align}
$C$ and $J$ being integration constants. 
By virtue of Eqs. (\ref{constant_1}) and (\ref{constant_2}), 
Eq. (\ref{Schwarzschild-geodetica-r})  reads as 
\begin{equation} \label{constant_3}
\dfrac{{\rm d} }{{\rm d}\lambda} \left[  A^{-1}(r(\lambda)) 
\left(
\left(\dfrac{{\rm d} r}{{\rm d}\lambda}\right)^2 +C^{2}\right) 
+\dfrac{J^2}{r^2}\right]=0, 
\end{equation}
and hence we arrive at 
\begin{equation}\label{r-equation_1}
 A^{-1}(r(\lambda)) \left[ \left(\dfrac{{\rm d} r}{{\rm d}\lambda}\right)^2 
+C^{2}\right] +\dfrac{J^2}{r^2}= \mathcal{E}, 
\end{equation}
where $\mathcal{E}>0$ is a constant.

The squared line element evaluated via (\ref{Schwarzschild-metric-1}) and with $\theta=\pi /2$ 
reads as 
\begin{equation}
\left. {\rm d}s^2 \right \vert_{\theta=\pi/2} = A(r) {\rm d}\tau^2 
+  A^{-1}(r) {\rm d}r^2 + r^2 {\rm d}\phi^2,
\end{equation}
then from Eqs. (\ref{constant_1}), (\ref{constant_2}), and (\ref{r-equation_1}) we obtain the useful relation
\begin{equation}
{\rm d}s^2 = \mathcal{E} {\rm d}\lambda^2,
\end{equation}
which makes it possible to write the equations defining geodesic motion as
\begin{subequations}
\label{Schwarzschild-geod-eqs-1}
\begin{align}
\left( \dfrac{{\rm d}r }{{\rm d}s} \right)^2  &=\left(1-\dfrac{2M}{r}\right)
\left(1-\dfrac{L^2}{r^2}\right) - C^{2} E^2,
\label{Schwarzschild-geod-eqs-1-variable-r}
\\
\dfrac{{\rm d}\phi}{{\rm d}s} &=\dfrac{L}{r^2},
\label{Schwarzschild-geod-eqs-1-variable-phi}
\\
\dfrac{{\rm d}\tau}{{\rm d}s} &=\dfrac{CE}{\left(1-\dfrac{2M}{r}\right)},
\label{Schwarzschild-geod-eqs-1-variable-tau}
\end{align}
\end{subequations}
where we have defined the real-valued constants $E$ and $L$ as
\begin{subequations}
\begin{align}
E &\equiv \dfrac{1}{\sqrt{\mathcal{E}}}, 
\label{E-definition}
\\ 
L &\equiv \dfrac{J}{\sqrt{\mathcal{E}}}=JE.
\label{J-definition}
\end{align}
\end{subequations}

Upon introducing the variable
\begin{equation}
\label{variable-u-def}
u={1 \over r},
\end{equation}
Eq. (\ref{Schwarzschild-geod-eqs-1}) can be equivalently written as
\begin{align}
\left({{\rm d}u \over {\rm d}\phi}\right)^2 &= 
{1 \over L^{2}}\left({dr \over ds}\right)^{2}=
\mathcal{F}(u),
\label{Schwarzschild-geod-eqs-2}
\\
{{\rm d}s \over {\rm d}\phi} &={1 \over Lu^2},
\\
{{\rm d}\tau \over {\rm d}\phi} &={CE \over Lu^{2}(1-2Mu)},
\end{align}
where 
\begin{equation} \label{F(u)}
\mathcal{F}(u) \equiv 2Mu^3-u^2-\left(\dfrac{2M}{L^2}\right)u 
+ \left(\dfrac{1-C^{2}E^2}{L^2}\right)
=2M(u-u_1)(u-u_2)(u-u_3).
\end{equation}
The above differential equations completely determine the geodesic  
motion in Euclidean Schwarzschild geometry in the equatorial plane $\theta = \pi /2 $. 
The turning points are described by the cubic equation
\begin{equation} \label{cubic-equation-F(u)=0}
\mathcal{F}(u)  =0,
\end{equation}
whose roots, say $u_1$, $u_2$ and $u_3$, satisfy the following equalities (Vi\`{e}te's formulae):
\begin{equation}
u_1+u_2+u_3 ={1 \over 2M},
\label{u1+u2+u3}
\end{equation}
\begin{equation}
u_1 u_2 + u_2 u_3 + u_3 u_1 =-{1 \over L^2},
\label{u1u2 + u1u3 + u2u3}
\end{equation}
\begin{equation}
u_1 u_2 u_3 = -{\left(1-C^{2}E^2\right) \over 2ML^2}.
\label{u1u2u3}
\end{equation}

\section{Roots of the cubic equation $\mathcal{F}(u)=0$. Qualitative analysis of the orbits}\label{Sec:roots-of-cubic}

The cubic equation $\mathcal{F}(u)=0$ can be re-expressed in 
canonical form \cite{Franci1979,Zwillinger2003}
\begin{equation}
w^{3}+pw+q=0,
\label{canonical-cubic}
\end{equation}
where
\begin{align}
p&=-\left({1 \over L^{2}}+{1 \over 12M^{2}}\right),
\label{canonic-p}
\\
q&=-{1 \over 108M^{3}}+{1 \over 6ML^{2}}(2-3C^2E^{2}).
\label{canonic-q}
\end{align}
Hence the discriminant $\bigtriangleup$ is given by
\begin{eqnarray}
\bigtriangleup &=& -(4p^{3}+27q^{2})
\nonumber \\
&=& {1 \over 4 (ML)^{2}}
\left[16 \left({M \over L}\right)^{4}
-(27C^4E^{4}-36C^2E^{2}+8)\left({M \over L}\right)^{2}
+(1-C^2E^{2})\right].
\label{Delta-cubic-expression}
\end{eqnarray}
From the above equations, it is clear that the integration constant $C$ (cf. Eq. \eqref{constant_1}) is a multiplicative constant  and hence can be set to one without loss of generality. However, in order to keep our analysis as general as possible, we here continue  employing a generic $C$.

\begin{figure*}[t!]
    \centering
    \includegraphics[scale=0.72]{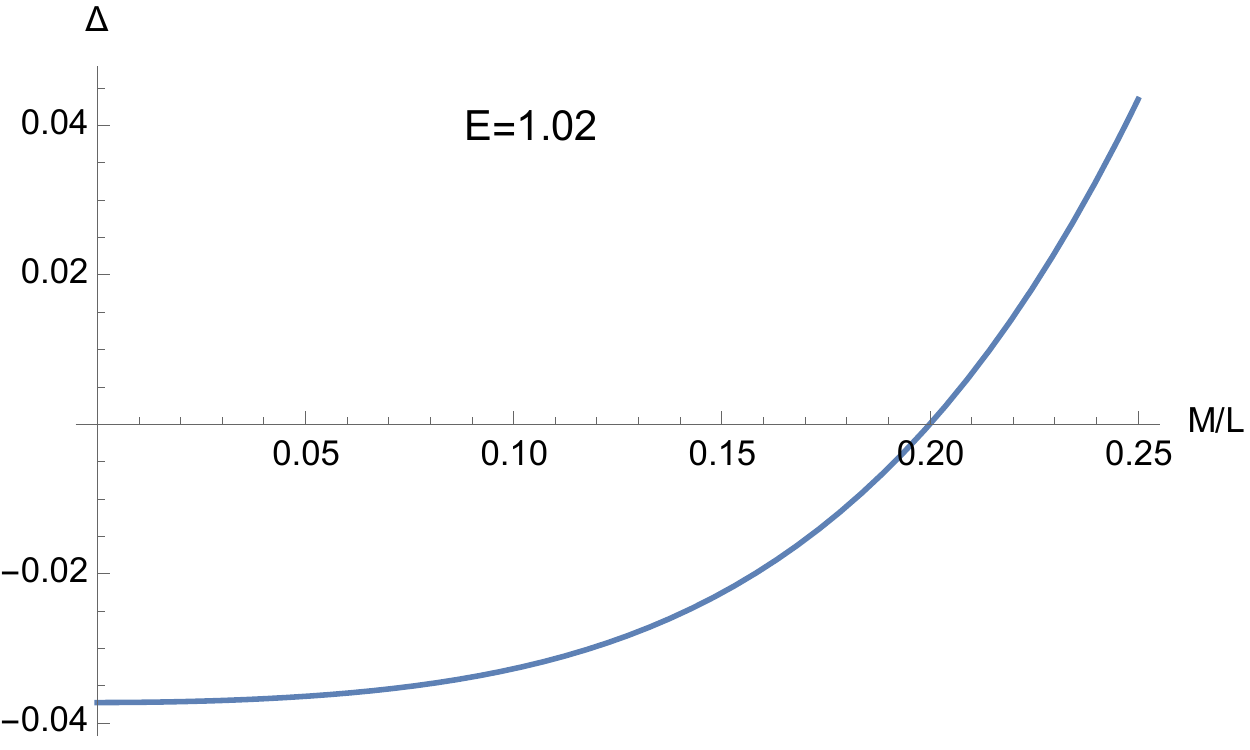}
    \caption{The discriminant (\ref{Delta-cubic-expression}) obtained with $E=1.02$ and $C=1$. 
It is clear that $\Delta$ assumes either positive, negative, or vanishing values.}
    \label{fig:discriminant-1}
\end{figure*} 
\begin{figure*}[t!]
    \centering
    \includegraphics[scale=0.42]{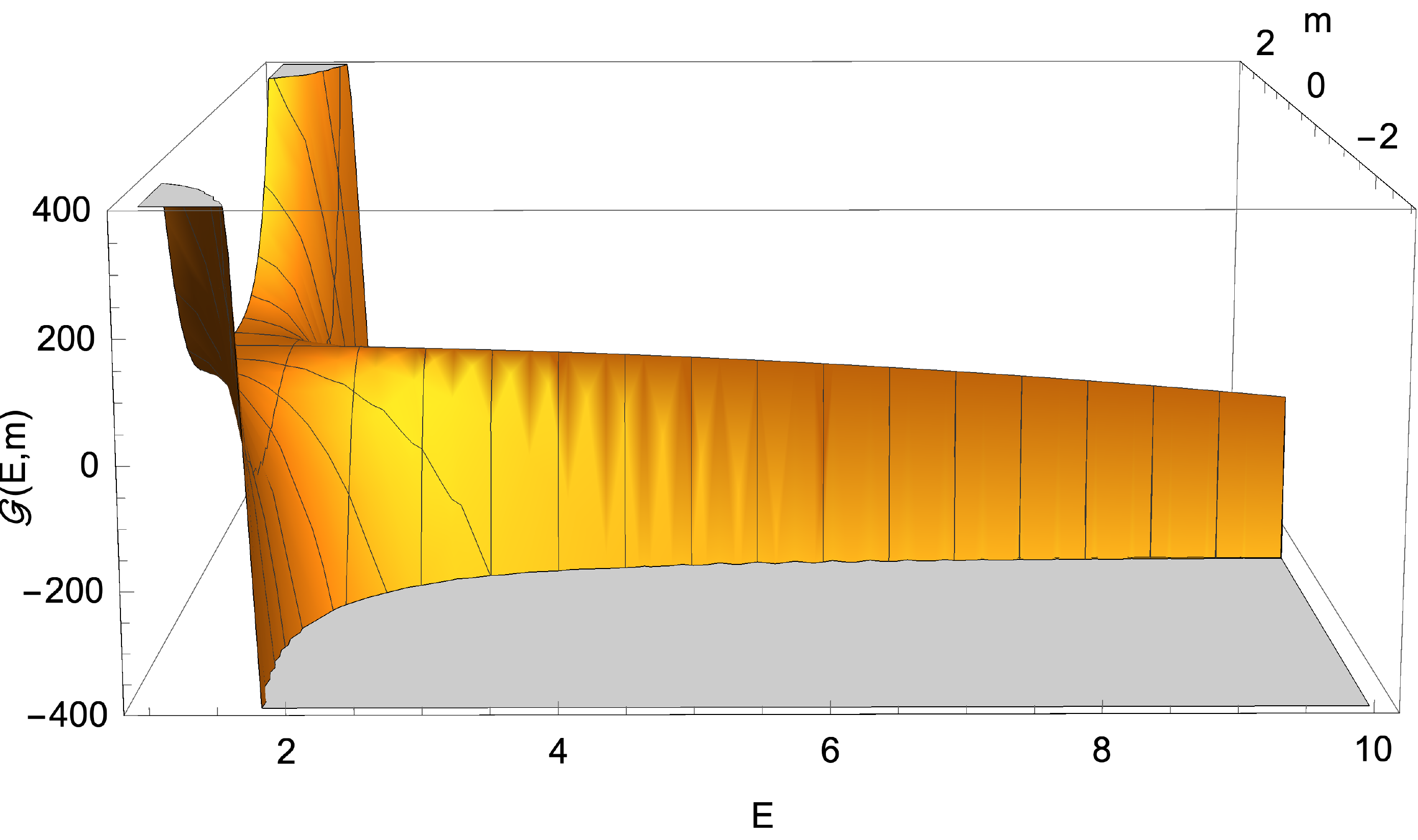}
    \caption{The function (\ref{mathscr-G(E,m)}) with $C=1$. It is clear that   
the discriminant (\ref{Delta-cubic-expression})  can be either positive, negative, or zero  
if $C^2E^2 >1$ (cf. Eq. (\ref{Delta-and-mathscr-G(E,m)})).}
    \label{fig:discriminant-3D-1}
\end{figure*} 
\begin{figure*}[t!]
    \centering
    \includegraphics[scale=0.72]{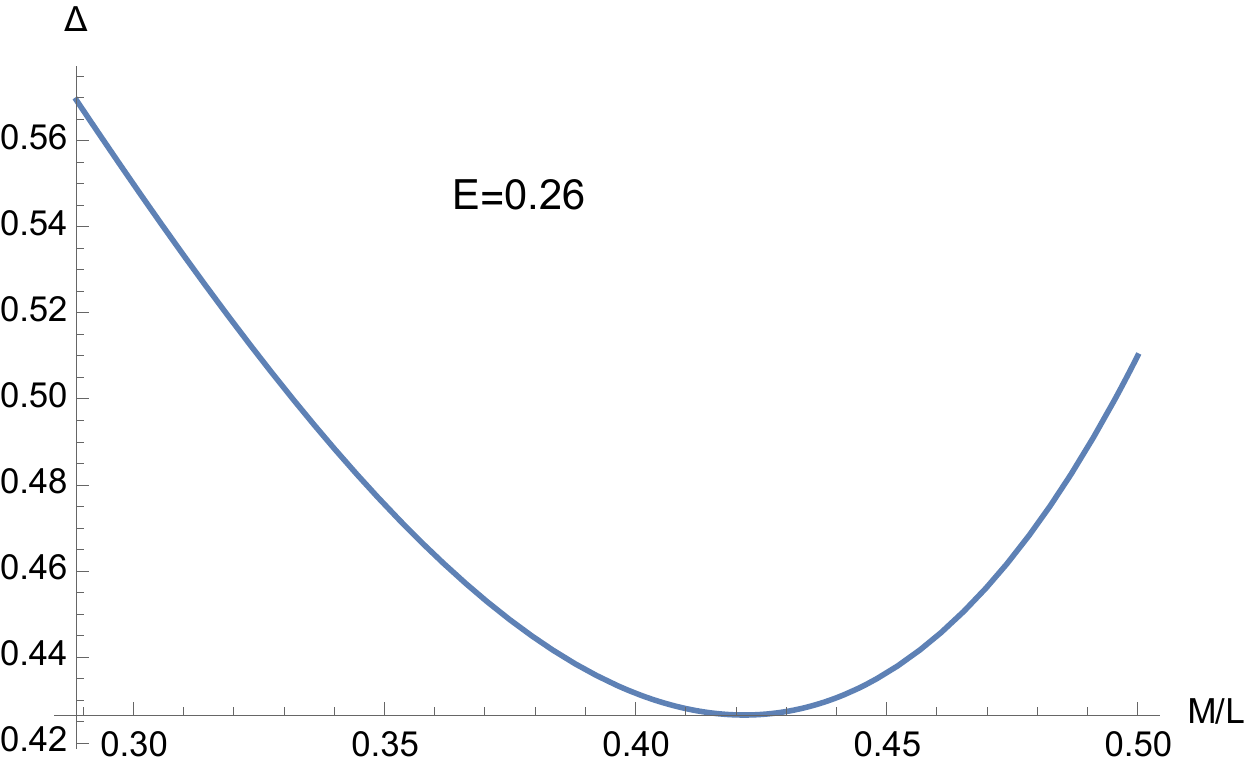}
    \caption{The discriminant (\ref{Delta-cubic-expression}) obtained with $E=0.26$ and $C=1$. 
It is clear that $\Delta$ never becomes negative.}
    \label{fig:discriminant-2}
\end{figure*} 
\begin{figure*}[t!]
    \centering
    \includegraphics[scale=0.42]{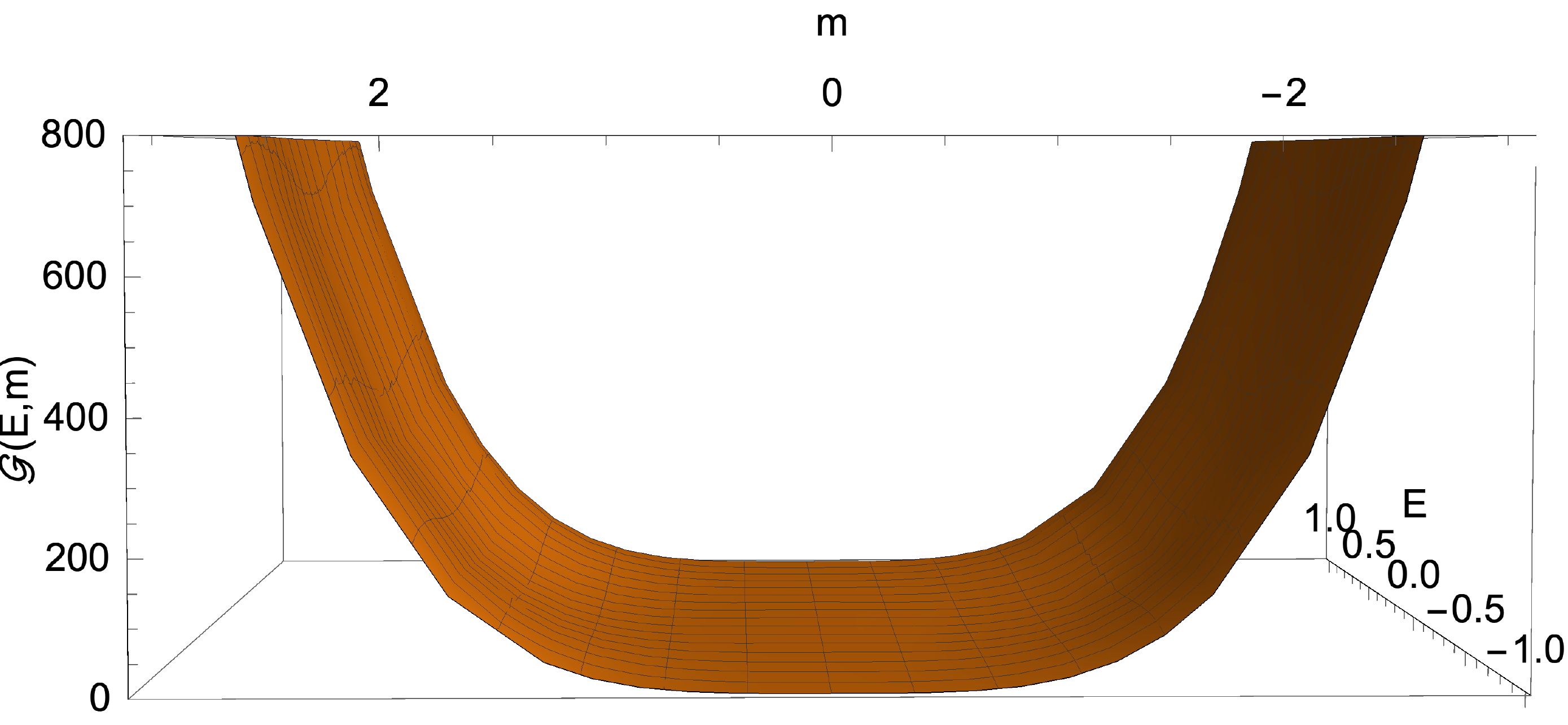}
    \caption{The function (\ref{mathscr-G(E,m)}) with $C=1$. It is clear that   
the discriminant (\ref{Delta-cubic-expression}) never becomes negative provided 
$C^2E^2 \leq1$ (cf. Eq. (\ref{Delta-and-mathscr-G(E,m)})).}
    \label{fig:discriminant-3D-2}
\end{figure*} 
\begin{figure}[t!]
         \centering
         \includegraphics[scale=0.72]{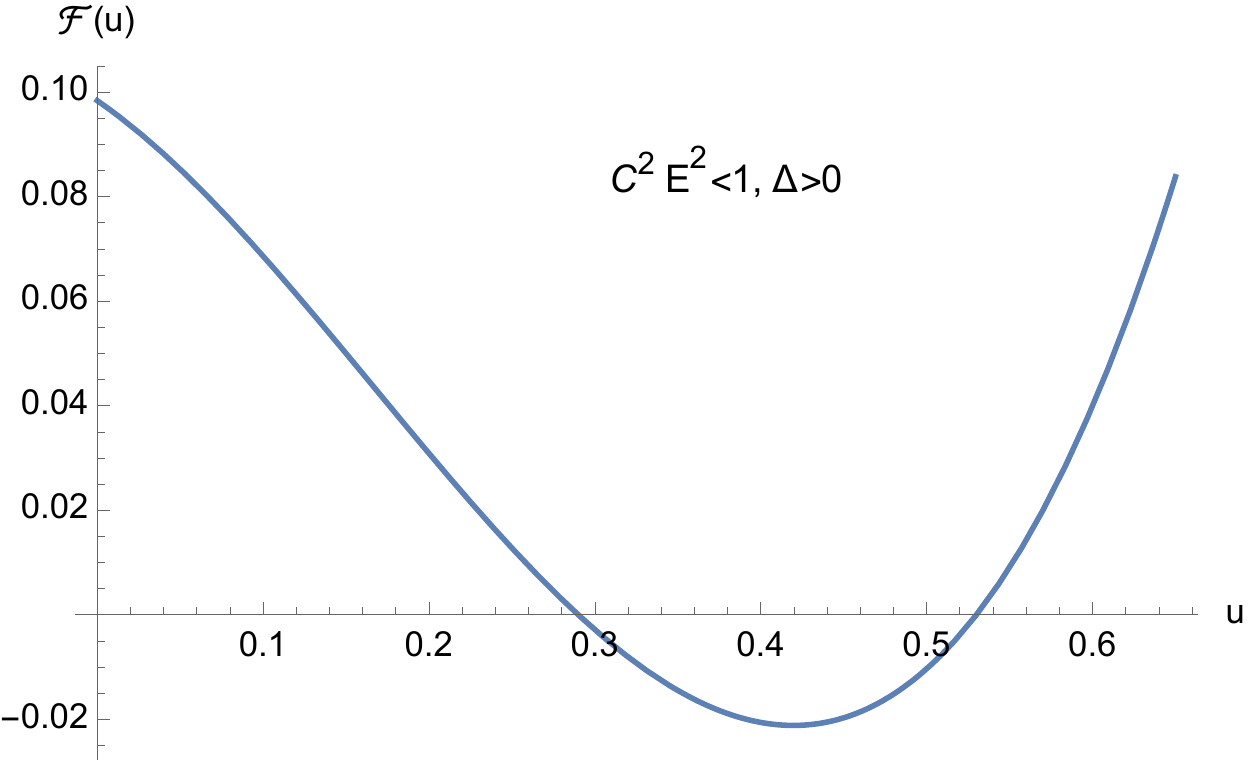}
         \caption{The positive roots of Eq. (\ref{cubic-equation-F(u)=0}) when  $C^2E^2<1$ and $\Delta>0$.}
         \label{fig:plot-f(u)-en-less-1-Delta-positive}
     \end{figure}
     \begin{figure}[t!]
         \centering
         \includegraphics[scale=0.72]{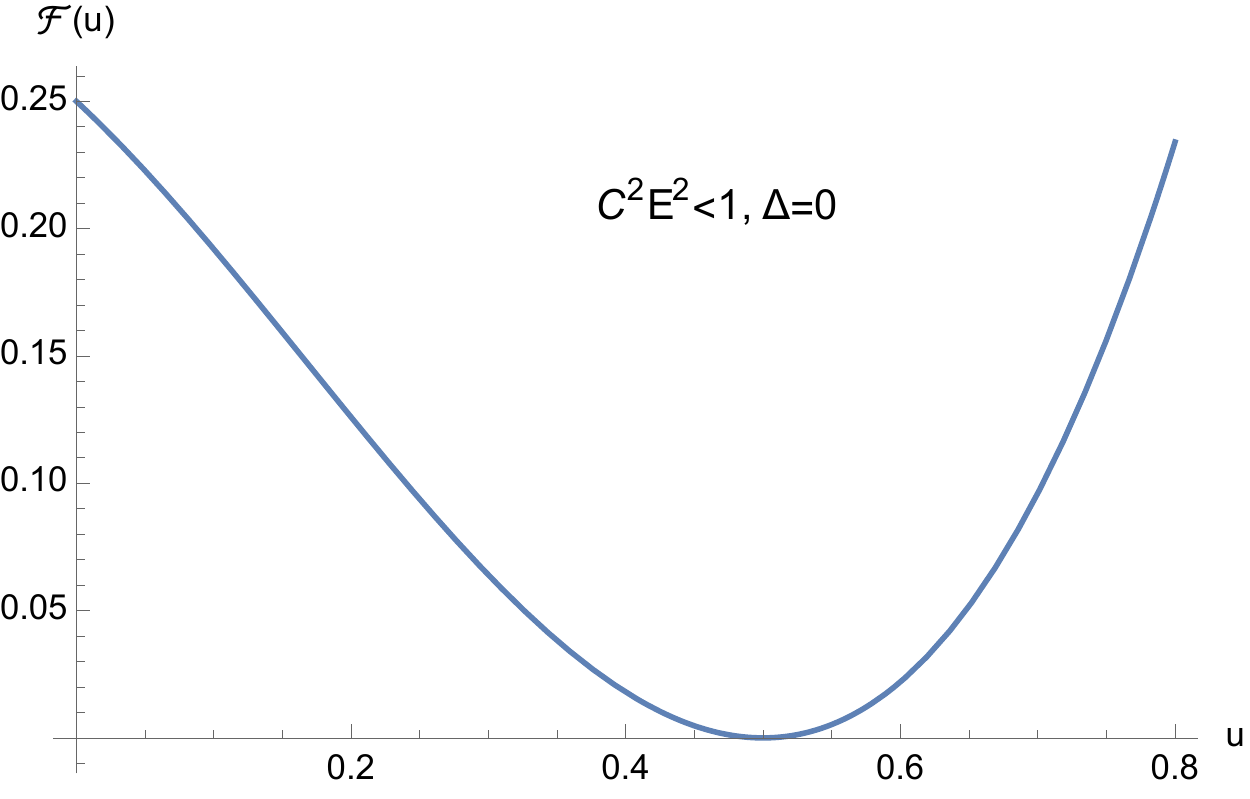}
         \caption{The positive roots of Eq. (\ref{cubic-equation-F(u)=0}) when  $C^2E^2<1$ and $\Delta=0$.}
         \label{fig:plot-f(u)-en-less-1-Delta-zero}
     \end{figure}
   \begin{figure}[t!]
         \centering
         \includegraphics[scale=0.72]{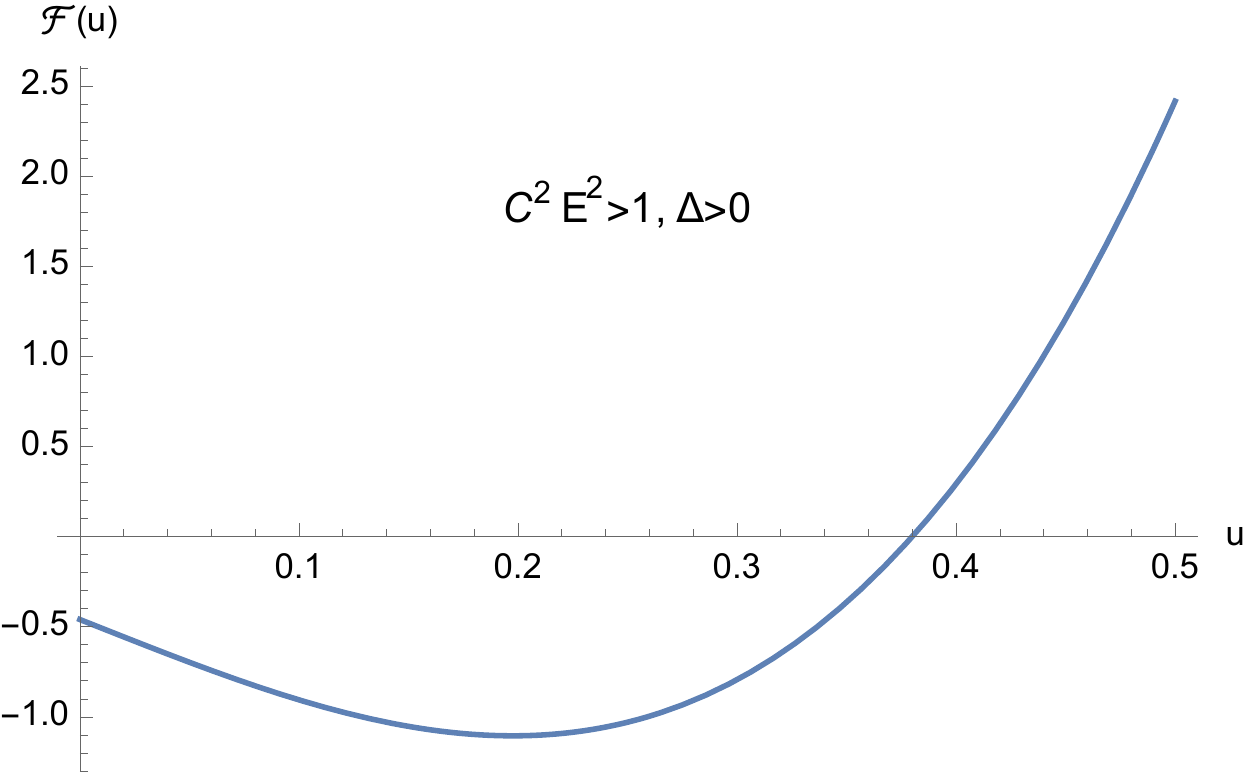}
         \caption{The positive root of Eq. (\ref{cubic-equation-F(u)=0}) when  $C^2E^2>1$ and $\Delta>0$.}
         \label{fig:plot-f(u)-en-greater-1-Delta-positive}
     \end{figure}
     \hfill
     \begin{figure}[t!]
         \centering
         \includegraphics[scale=0.72]{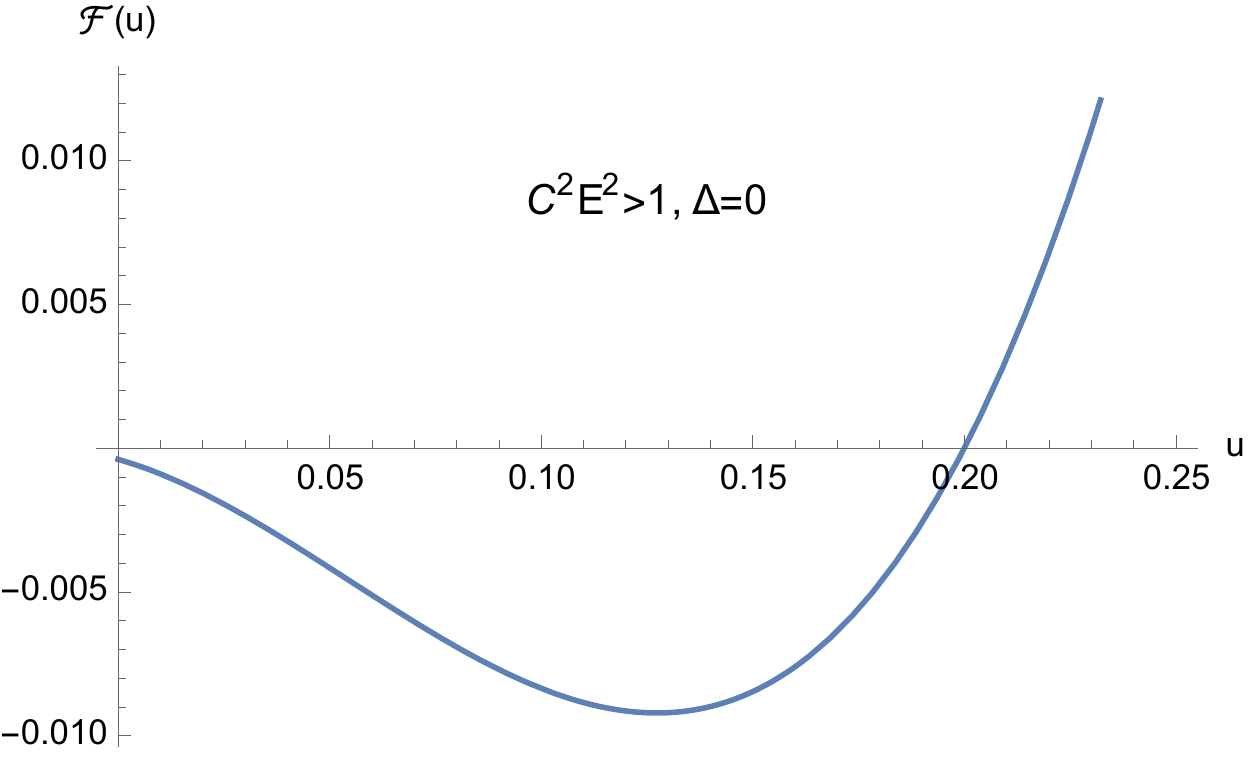}
         \caption{The positive root Eq. (\ref{cubic-equation-F(u)=0}) when $C^2E^2>1$ and $\Delta=0$.}
         \label{fig:plot-f(u)-en-greater-1-Delta-zero}
     \end{figure}
      \hfill
     \begin{figure}[t!]
         \centering
         \includegraphics[scale=0.72]{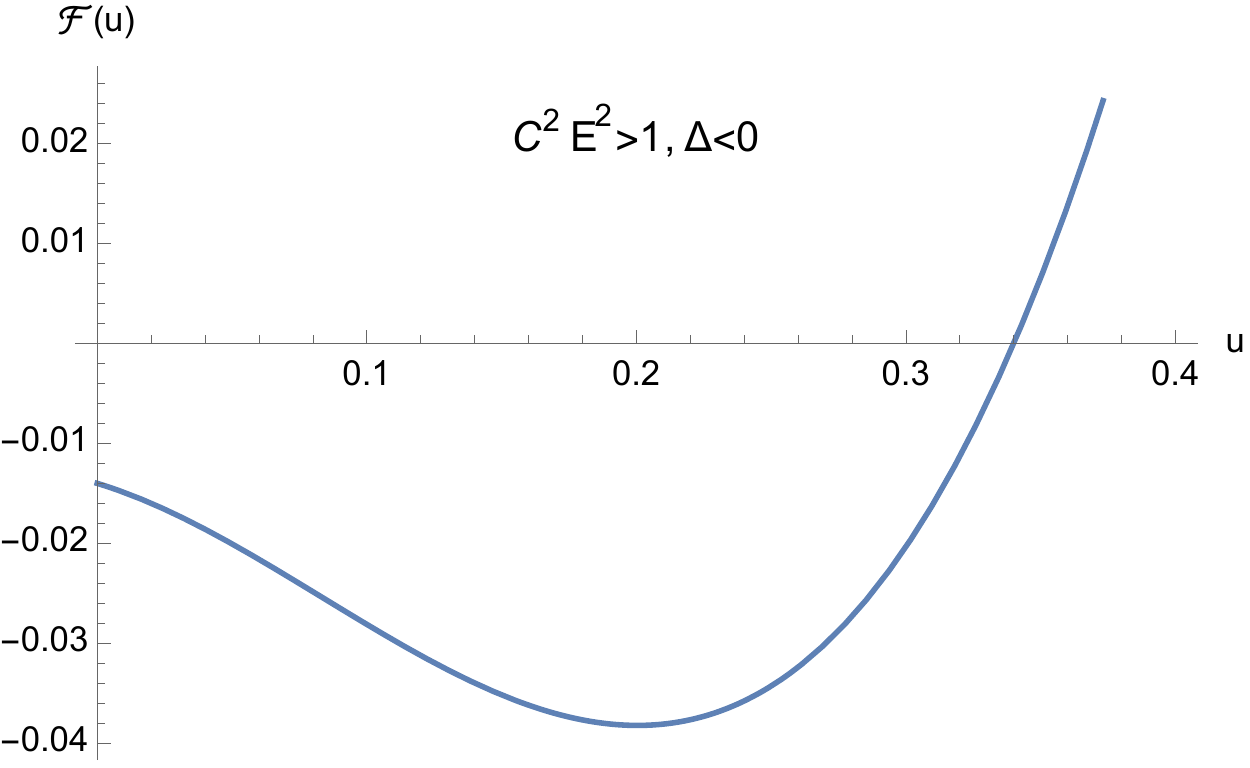}
         \caption{The positive root of Eq. (\ref{cubic-equation-F(u)=0})  when $C^2E^2>1$ and $\Delta<0$.}
         \label{fig:plot-f(u)-en-greater-1-Delta-negative}
     \end{figure}
\begin{figure}[t!]
         \centering
         \includegraphics[scale=0.72]{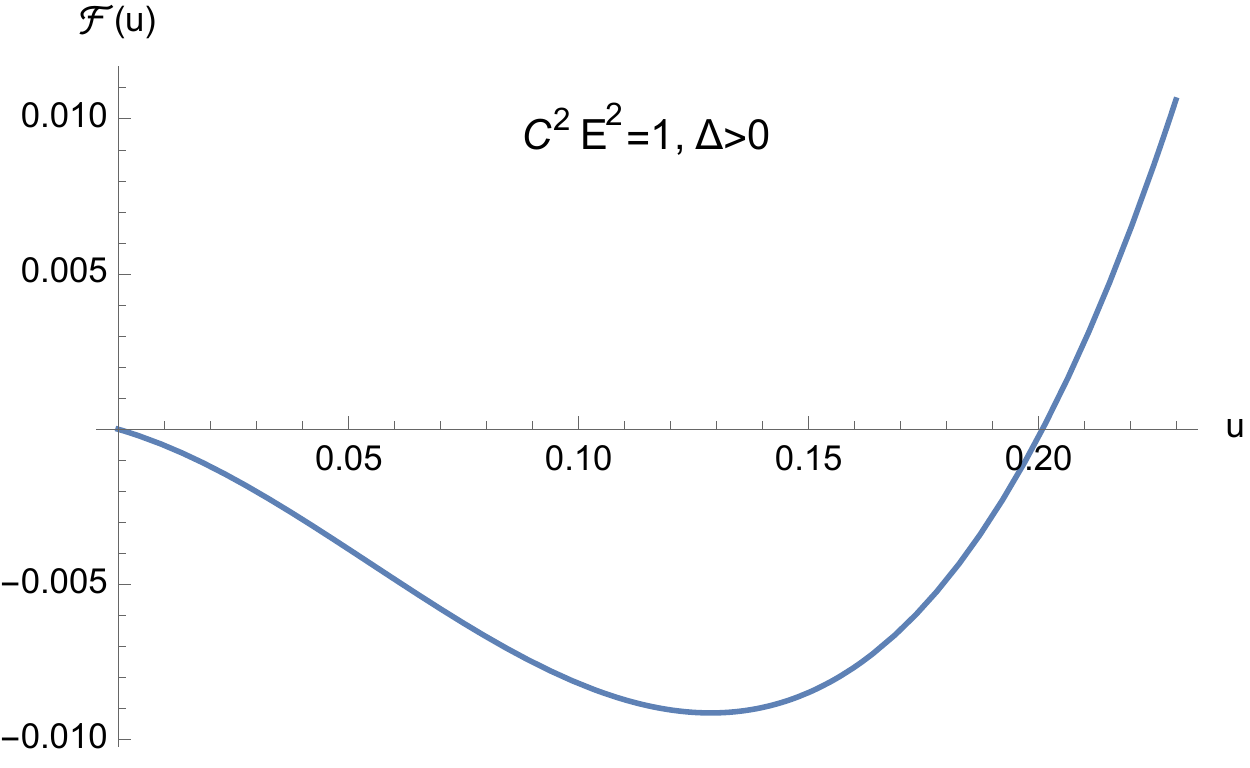}
         \caption{The non-negative roots of Eq. (\ref{cubic-equation-F(u)=0}) when $C^2E^2=1$ and $\Delta>0$.}
         \label{fig:plot-f(u)-en=1-Delta-positive}
     \end{figure}
     \hfill
     \begin{figure}[t!]
         \centering
         \includegraphics[scale=0.72]{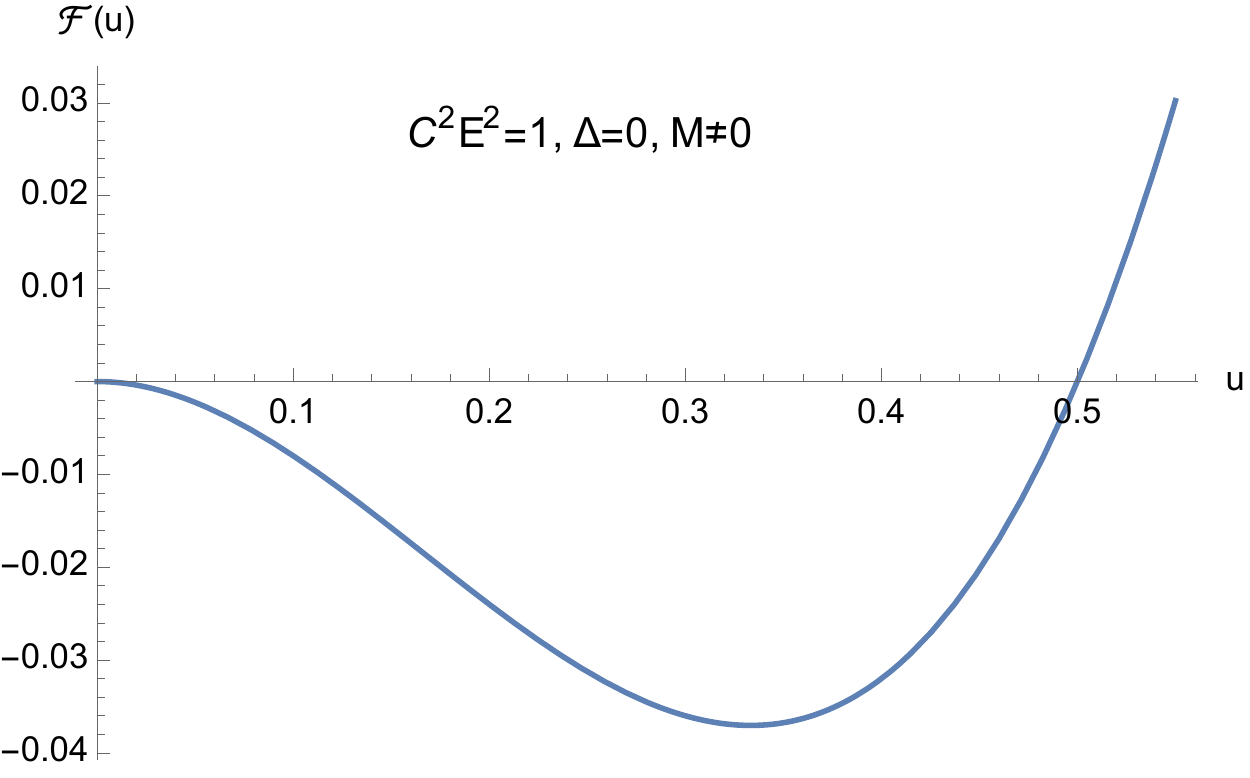}
         \caption{The roots of Eq. (\ref{cubic-equation-F(u)=0}) when $C^2E^2=1$, $\Delta=0$, and $M \neq 0$.}
         \label{fig:plot-f(u)-en=1-Delta-zero}
     \end{figure}

The sign of  $\Delta$
depends on the behaviour of the real-valued function
\begin{equation}\label{mathscr-G(E,m)}
\mathscr{G} \left(E,m\right)=16m^4 -\left(27C^4E^4-36C^2E^2+8\right)m^2 +\left(1-C^2E^2\right),
\end{equation}
where
\begin{equation}
m \equiv \dfrac{M}{L},
\end{equation}
the discriminant (\ref{Delta-cubic-expression}) being in fact expressible as
\begin{equation}\label{Delta-and-mathscr-G(E,m)}
\Delta =\dfrac{1}{4\left(M^2 L\right)^2} \mathscr{G} \left(E,m\right).
\end{equation}
In this way, we find that $\Delta$ can be either positive, negative, 
or zero only if $C^2E^2 >1$ (see Figs. \ref{fig:discriminant-1} and \ref{fig:discriminant-3D-1}), 
whereas when $C^2E^2 \leq 1$ we only have $\Delta \geq 0$ (see Figs. \ref{fig:discriminant-2} 
and \ref{fig:discriminant-3D-2}). In particular, in this second case, $\Delta=0$ if
\begin{subequations}
\label{Delta=0,m=1/2,E=0}
\begin{align}
 \vert m \vert =\dfrac{1}{2},
 \label{Delta=0,m=1/2}
\\  
E=0,
\label{Delta=0,E=0}
 \end{align}
\end{subequations}
or
\begin{subequations}
\label{Delta=0,m=0,E=1}
\begin{align}
m=0, 
\\ 
 CE=1.
\end{align}
\end{subequations}
This means that the cubic (\ref{cubic-equation-F(u)=0}) can only admit real roots as soon as $C^2E^2 \leq 1$.  
This is different from the Lorentzian case, where complex roots 
can arise both with $C^2E^2 >1$ and $C^2E^2 \leq 1$.

From the theory of cubic equations it is known that multiple roots arise when the discriminant 
(\ref{Delta-cubic-expression}) vanishes. 
In particular, if $\Delta =0$ and $p=0$, $w_1=w_2=w_3=0$ is a triple root of 
(\ref{canonical-cubic}). On the other hand, if  $\Delta =0$ and $p \neq 0$, then 
$w_1= 3q/p$ is a single root, while $w_2=w_3 = -3q/(2p)$ is a double root of  
(\ref{canonical-cubic}). In our case, from Eq. (\ref{canonic-p}) it is clear that  
$\Delta$ and $p$ cannot vanish simultaneously. This means that the cubic equation 
(\ref{cubic-equation-F(u)=0}) never admits a triple root when (\ref{Delta-cubic-expression}) 
vanishes. Furthermore, by employing Descartes' rule of signs and bearing in mind the 
discriminant   (\ref{Delta-cubic-expression}), we have the following situation:
\begin{itemize}
\item $C^2E^2<1$:
\begin{itemize}
\item[ i)]  $\Delta >0$. The cubic  (\ref{cubic-equation-F(u)=0}) has two   
distinct positive roots and one negative root (see Fig. \ref{fig:plot-f(u)-en-less-1-Delta-positive});
\item[ ii)] $\Delta =0$. The cubic  (\ref{cubic-equation-F(u)=0}) admits one 
negative root and two coincident positive roots  (see Fig. 
\ref{fig:plot-f(u)-en-less-1-Delta-zero} and Eq. (\ref{Delta=0,m=1/2,E=0})) which read as 
\begin{subequations}
\label{u1,u2,u3-with-E^2<1-Delta-zero}
\begin{align}
u_1 &=-\dfrac{1}{2M}, 
\\
u_2=u_3 &=\dfrac{1}{2M},
\label{u2-u3-coincide}
\end{align}
\end{subequations}
respectively.
\end{itemize}
\item $C^2E^2>1$:
\begin{itemize}
\item[ i)] $\Delta >0$. The cubic (\ref{cubic-equation-F(u)=0}) presents  
one positive root and two distinct negative roots (see Fig. \ref{fig:plot-f(u)-en-greater-1-Delta-positive});
\item[ ii)] $\Delta =0$. The cubic (\ref{cubic-equation-F(u)=0}) admits one 
positive root and two coincident negative roots (see Fig. \ref{fig:plot-f(u)-en-greater-1-Delta-zero});
\item[ iii)] $\Delta <0$. The cubic (\ref{cubic-equation-F(u)=0}) exhibits one 
positive root and two complex conjugate roots (see Fig. \ref{fig:plot-f(u)-en-greater-1-Delta-negative}). 
\end{itemize}
\item $C^2E^2 =1$:
\begin{itemize}
\item[ i)] $\Delta >0$. The cubic (\ref{cubic-equation-F(u)=0}) has 
one vanishing root, the negative root 
\begin{equation}\label{E=1-root-negative}
u_1=\dfrac{1-\sqrt{1+16\dfrac{M^2}{L^2}}}{4M}
\end{equation}
and the positive root 
\begin{equation}\label{E=1-root-positive}
u_2=\dfrac{1+\sqrt{1+16\dfrac{M^2}{L^2}}}{4M},
\end{equation}
see Fig. \ref{fig:plot-f(u)-en=1-Delta-positive};
\item[ ii)] $\Delta =0$ with $M \neq 0$ and $\vert L\vert  \gg M$ (see Eq. (\ref{Delta=0,m=0,E=1})). 
By virtue of Eqs. (\ref{E=1-root-negative}) and (\ref{E=1-root-positive}), 
we find that the cubic (\ref{cubic-equation-F(u)=0}) admits a vanishing root 
(with multiplicity two) and the positive root $u_2=\dfrac{1}{2M}$ 
(see Fig. \ref{fig:plot-f(u)-en=1-Delta-zero}).
\end{itemize}
\end{itemize}

From Figs. \ref{fig:plot-f(u)-en-less-1-Delta-positive}--\ref{fig:plot-f(u)-en=1-Delta-zero} 
it is clear that the conditions
\begin{align*}
 & 0<u_1<u<u_2, 
 \nonumber \\
 & \mathcal{F}(u) > 0,
\end{align*} 
never hold simultaneously. This is due to the fact that when (\ref{cubic-equation-F(u)=0}) 
admits two positive roots (i.e., when $C^2E^2<1$) the function (\ref{F(u)}) is such that 
$\mathcal{F}(0) >0$. As a consequence, no elliptic-like orbits exist in Euclidean 
Schwarzschild geometry, unlike the corresponding Lorentzian pattern. Furthermore, since 
$\mathcal{F}(0) >0$ only if $C^2E^2<1$, we have the following classification:
\begin{align}
C^2E^2 & <1: \quad {\rm unbounded \; orbits},
\label{E^2<1-orbits-classification}
 \\
C^2E^2 & >1: \quad {\rm  bounded \;orbits},
\label{E^2>1-orbits-classification}
\end{align}
which amounts to the reversed situation with respect to general relativity. Here, bounded (resp. unbounded) orbits are defined as those trajectories where $r$ remains bounded (resp. unbounded).

From  Eq. (\ref{Schwarzschild-geod-eqs-1-variable-r}) we see that   
$\left({\rm d}r/{\rm d}s\right)^2 <0$  if $r=2M$. Therefore, the condition 
(\ref{r-bigger-2M}) should be tightened and for this purpose we impose
\begin{equation} \label{r-bigger-2M-2}
r>2M.
\end{equation}
In light of the above condition, the lower bound
\begin{equation}
r > \vert L \vert
\end{equation}
is a necessary but not sufficient condition to ensure that 
$\left({\rm d}r/{\rm d}s \right)^2 >0$. 

Hereafter, we will limit our analysis to geodesics enforcing the 
constraint
\begin{align} 
 u &< \dfrac{1}{2M}, \label{u<1/2M}
\end{align}
jointly with $\mathcal{F}(u) \geq 0$ (see Eq. \eqref{Schwarzschild-geod-eqs-2}).

\section{Solution in terms of elliptic integrals}\label{Sec:Sol-elliptic.intregrals}

As we have shown before, the algebraic equation 
of third degree (\ref{cubic-equation-F(u)=0}) involves three real roots as soon as 
$C^2E^2 \leq 1$. In particular, when $C^2E^2<1$  and the discriminant 
(\ref{Delta-and-mathscr-G(E,m)}) is non-vanishing, the solution $u_3$  
turns out to admit the lower bound (see Appendix \ref{Appendix:general-formulae-roots} 
for further details) 
\begin{equation}
u_3 \geq  \dfrac{1}{2M},
\end{equation}
with $u_3=1/(2M)$ in the case $E=0$. On the other hand, Figs. 
\ref{fig:plot-f(u)-en-less-1-Delta-positive} and 
\ref{fig:plot-f(u)-en-less-1-Delta-zero} clearly indicate that  
the case $C^2E^2<1$ could, in principle, entail both first-kind trajectories, for which 
$0< u\leq u_2$, and second-kind ones, where $u>u_3 $ (this is our definition of first-kind and second-kind orbits). Since the latter neither 
obey (\ref{r-bigger-2M-2}) nor belong to the real section of the complexified 
Schwarzschild spacetime, our calculations will be restricted to first-kind orbits. 
This represents a clear difference with respect to general relativity, where second-kind trajectories 
are always allowed. 

Under the hypothesis $0<C^2E^2<1$, the three real solutions of the cubic 
(\ref{cubic-equation-F(u)=0}) can be parametrised as
\begin{subequations}
\label{u1,u2,u3-E^2<1}
\begin{align}
u_1 &= -\dfrac{1}{\ell} \left( e-1\right), 
\label{u1-E^2<1}
\\
u_2 &= \dfrac{1}{2M} -\dfrac{2}{\ell}, 
\label{u2-E^2<1}
\\
u_3 &= \dfrac{1}{\ell} \left( e+1\right),
\label{u3-E^2<1}
\end{align}
\end{subequations}
where we have adopted a choice which does not resemble exactly the Lorentzian-signature 
framework \cite{Chandrasekhar1983} (see Appendix \ref{Sec:roots-cubic-E^2<1} for details). 

The roots (\ref{u1,u2,u3-E^2<1}) clearly satisfy Eq. (\ref{u1+u2+u3}) and in addition
\begin{equation} \label{u1,u,2u3-E^2<1-inequalities}
u_1 <0 <u_2 <\dfrac{1}{2M}< u_3,
\end{equation}
provided that
\begin{subequations}
\label{constraints_1}
\begin{align}
\ell & >0, 
\label{l>0}
 \\
e & >1,
\label{e>1}
 \\
\dfrac{1}{2\left(e+1\right)}<\mu &<\dfrac{1}{4},
\label{mu<1/4}
\end{align}
\end{subequations}
where we have defined 
\begin{equation}\label{mu-definition}
\mu \equiv \dfrac{M}{\ell}.
\end{equation}
It follows from Eqs. \eqref{l>0} and \eqref{e>1} that, similarly to the 
Lorentzian-signature pattern, we can interpret the positive constant 
$\ell$ as the \emph{latus rectum} and $e$ as the eccentricity; indeed, 
we will see that our investigation predicts the existence of  
trajectories which display a formal analogy with the hyperbolic orbits of general relativity (see Figs. \ref{fig:E^2<1-first-kind-orbit} and 
\ref{fig:E^2<1-first-kind-orbit-limiting-case}, below).   

Vi\`{e}te's formulae  (\ref{u1u2 + u1u3 + u2u3}) and (\ref{u1u2u3}) yield 
\begin{subequations}
\label{relations-1/L^2-E^2<1}
\begin{align}
\dfrac{1}{L^2} &= \dfrac{\mu \left(3+e^2\right)-1}{M \ell},
\label{inequality-1/L^2}
\\
\dfrac{\left(1-C^2E^2\right)}{L^2} &= \dfrac{\left(e^2-1\right)\left(1-4 \mu \right)}{\ell^2},
\end{align}
\end{subequations}
respectively, and we recognize that the set of constraints (\ref{constraints_1}) guarantees also 
that
\begin{align}
L^2 &>0, 
\\
0 <C^2E^2 &<1.
\label{0<E^2<1}
\end{align}

In the hypothetical case 
\begin{equation} \label{mu-u2=u3}
\mu = \dfrac{1}{\left(6+2e\right)},
\end{equation}
the roots (\ref{u2-E^2<1}) and (\ref{u3-E^2<1}) would coincide and relations 
(\ref{relations-1/L^2-E^2<1}) would be turned into
\begin{subequations}
\label{parameters,u2=u3}
\begin{align}
\dfrac{L^2}{M^2} &=\dfrac{4 \left(3+e\right)^2}{\left(e+1\right)\left(e-3\right)},
\\
\left(1-C^2E^2\right) &=\dfrac{\left(e^2-1\right)}{\left(e^2-9\right)}.
\label{parameters,u2=u3-2}
\end{align}
\end{subequations}
However, Eqs. (\ref{u1,u2,u3-E^2<1})--(\ref{constraints_1}), as well as Eq. 
(\ref{0<E^2<1}), do not account for this scenario. Indeed, we know that when 
$u_2=u_3$ both (\ref{Delta=0,m=1/2,E=0}) and (\ref{u2-u3-coincide}) are satisfied, but    
the latter implies that the constraint (\ref{u<1/2M}) is violated, while, in light of 
the former, Eq. (\ref{parameters,u2=u3}) cannot be valid; furthermore, it is clear that   
(\ref{Delta=0,E=0}) cannot stem from Eq. (\ref{0<E^2<1}). Therefore, our analysis of 
first-kind trajectories naturally implies, on the one hand,
\begin{equation}
u_2 \neq u_3,
\end{equation}
while, on the other hand, it includes also the limiting situation 
\begin{equation}\label{limiting-case}
u_2 \to \dfrac{1}{2M} .
\end{equation}

First-kind orbits having $C^2E^2<1$ (i.e., unbounded, see Eq. (\ref{E^2<1-orbits-classification})) 
will be dealt with in the following section.

\subsection{First-kind orbits having $C^2 E^2 <1$}

As pointed out before, the case $C^2E^2<1$ consists of unbounded first-kind orbits only. This means that,
equivalently, our study will rely on one portion of Fig. 
\ref{fig:plot-f(u)-en-less-1-Delta-positive} only, whereas the situation depicted in Fig. 
\ref{fig:plot-f(u)-en-less-1-Delta-zero} will be ignored.

Orbits of first kind are constrained by means of 
\begin{equation} \label{u-range-first-kind}
0<u \leq u_2 <\dfrac{1}{2M},
\end{equation}
see Fig. \ref{fig:plot-f(u)-en-less-1-Delta-positive}.

Starting from Eqs. \eqref{variable-u-def}--\eqref{F(u)}, the system of differential equations for the geodesic motion can be re-expressed
in the form (where $\varepsilon=\pm 1$)  
\begin{align}
{\dd \tau \over \dd s}&= {\dd \tau \over \dd r}{\dd r \over \dd s}={CE \over \left(1-{2M \over r}\right)},
\label{(3.1)}
\\
{\dd r \over \dd s}&=\varepsilon \,\sqrt{\left(1-{2M \over r}\right)
\left(1-{L^{2}\over r^{2}}\right)
-C^{2}E^{2}},
\label{(3.2)}
\\
{\dd \phi \over \dd s}&={\dd \phi \over \dd r}{\dd r \over \dd s}={L \over r^{2}},
\label{(3.3)}
\end{align}
with the understanding that the physically relevant solution pertains 
to non-negative values of the argument of the square root on the right-hand
side of Eq. \eqref{(3.2)}. Moreover, as pointed out before, we focus on the case 
in which the root $u_{1}$ of the
equation $\mathcal{F}(u)=0$ is negative, while the roots $u_{2}$ and $u_{3}$
are positive and such that (cf. Eqs. \eqref{u1,u,2u3-E^2<1-inequalities} and   \eqref{u-range-first-kind})
\begin{equation}
u \leq u_{2} < u_{3}.
\label{(3.4)}
\end{equation}
We therefore find from Eqs.  \eqref{(3.1)}--\eqref{(3.3)}, upon setting 
$P_{3}(u)=\mathcal{F}(u)/(2M)=(u-u_{1})(u-u_{2})(u-u_{3})$, the following integral formulae
for the solution:
\begin{align}
s&=s_{0}+{\varepsilon \over L \sqrt{2M}} \int_{{1 \over r}}^{u_{2}}
{\dd u \over u^{2}\sqrt{P_{3}(u)}},
\label{(3.5)}
\\
\tau &=\tau_{0}+{\varepsilon CE \over L \sqrt{2M}}\int_{{1 \over r}}^{u_{2}}
{\dd u \over u^{2}(1-2Mu)\sqrt{P_{3}(u)}},
\label{(3.6)}
\\
\phi&=\phi_{0}+{\varepsilon \over \sqrt{2M}} \int_{{1 \over r}}^{u_{2}}
{\dd u \over \sqrt{P_{3}(u)}}.
\label{(3.7)}
\end{align}
Note that, in agreement with what we said before, the upper limit of
integration is $u_{2}$, in order to avoid negative values of
$P_{3}(u)$, which are unphysical. At this stage, it is convenient 
to apply twice the method of adding and subtracting $2Mu$ in the 
numerator of the integrand in Eq. \eqref{(3.6)}. Thus, upon defining
(our $n=0,1,2$)
\begin{subequations}
\begin{align}
J_{n}&=\int_{{1 \over r}}^{u_{2}}
{\dd u \over u^{n}\sqrt{P_{3}(u)}},
\label{(3.8)}
\\
I&=\int_{{1 \over r}}^{u_{2}}
{\dd u \over \left(u-{1 \over 2M}\right)\sqrt{P_{3}(u)}},
\label{(3.9)}
\end{align}
\end{subequations}
we obtain eventually the desired solution in the form
\begin{align}
s&=s_{0}+{\varepsilon \over L \sqrt{2M}}J_{2},
\label{(3.10)}
\\
\tau&=\tau_{0}+ CE(s-s_{0})
+\varepsilon CE{\sqrt{2M}\over L}(J_{1}-I),
\label{(3.11)}
\\
\phi&=\phi_{0}+{\varepsilon \over \sqrt{2M}}J_{0}.
\label{(3.12)}
\end{align}
The four integrals occurring in the solution \eqref{(3.10)}--\eqref{(3.12)}  
can be evaluated by means of incomplete elliptic integrals
(see Appendix \ref{Appendix:Incompl-integr}) according to the formulae \cite{Byrd}
\begin{subequations}
\begin{align}
a &=u_{3}, \quad b=u_{2}, \quad c=u_{1},
\label{(3.13)}
\\
\varphi &={\rm arcsin} \sqrt{(a-c)\left(b-{1 \over r}\right)
\over (b-c)\left(a-{1 \over r}\right)},
\label{(3.14)}
\\
k^{2} &={(b-c)\over (a-c)},
\label{(3.15)}
\\
\alpha^{2}&={a \over b}k^{2},
\label{(3.17)}
\\
\beta &=k \, \sqrt{{\left({1 \over 2M}-a \right)
\over \left({1 \over 2M}-b \right)}},
\label{(3.20)}
\end{align}
\end{subequations}
\begin{subequations}
\begin{align}
J_{0} &={2 \over \sqrt{a-c}} F(\varphi,k^{2}),
\label{(3.16)}
\\
J_{1}&={2 \over a \sqrt{a-c}}
\left[F(\varphi,k^{2})+\left({\alpha^{2}\over k^{2}}-1 \right)
\pi(\varphi,\alpha^{2},k^{2})\right],
\label{(3.18)}
\\
J_{2}&= {2 \over a^{2}\sqrt{a-c}} \left \{
F(\varphi,k^{2})+2\left({\alpha^{2}\over k^{2}}-1 \right)
\pi(\varphi,\alpha^{2},k^{2}) \right .
\nonumber \\
&+ \left({\alpha^{2}\over k^{2}}-1 \right)^{2}
{1 \over 2 (\alpha^{2}-1)(k^{2}-\alpha^{2})}
\Bigr[\alpha^{2}E(\varphi,k^{2}) 
\nonumber \\
&+ (k^{2}-\alpha^{2})F(\varphi,k^{2})
+(2 \alpha^{2}k^{2}+2\alpha^{2}-\alpha^{4}-3k^{2})
\pi(\varphi,\alpha^{2},k^{2})
\nonumber \\
&- \left . {\alpha^{4}{\rm sn}(u){\rm cn}(u){\rm dn}(u) 
\over (1-\alpha^{2}{\rm sn}^{2}(u))} \Bigr]
\right \},
\label{(3.19)}
\\
I&=-{2 \over (2M-a)\sqrt{a-c}}
\left[F(\varphi,k^{2})+\left({\beta^2 \over k^{2}}-1 \right)
\pi(\varphi,\beta^2,k^{2})\right].
\label{(3.21)}
\end{align}
\end{subequations}

At a deeper level, the solution of Eq. \eqref{(3.7)} for 
${1 \over r}=u(\phi)$ should not depend on the integration path. 
If one denotes by $\gamma$ a closed integration path and if one sets
\begin{equation}
{1 \over \sqrt{2M}} \int_{\gamma}{\dd u \over \sqrt{P_{3}(u)}}=\omega,
\label{(3.22)}
\end{equation}
this means that \cite{Hackmann2008}
\begin{equation}
\phi-\phi_{0}-\omega=\frac{1}{\sqrt{2M}}\int_{u}^{u_{2}}
{\dd u^\prime \over \sqrt{P_{3}(u^\prime)}},
\label{(3.23)}
\end{equation}
should hold as well. In other words, the desired solution should be
periodic of period $\omega$. At this stage, Eq. \eqref{(3.7)} is viewed as
defined on the Riemann surface of the algebraic function
$u \rightarrow \sqrt{P_{3}(u)}$. At the deep level of  complex analysis and algebraic geometry, this is the appropriate concept of periodicity \cite{Hackmann2008}, which should not be confused with the periodicity of the function $y=\cos \left(\tfrac{\tau}{4M}\right)\sqrt{\tfrac{r}{2M}-1} \,  {\rm exp}\left(\tfrac{r}{4M}\right)$ in Kruskal-Szekeres coordinates \cite{GH1977}.

\subsection{Graphical representation of unbounded first-kind orbits }

Having obtained the general  solution \eqref{(3.10)}--\eqref{(3.12)} of first-kind orbits that satisfy $C^2E^2<1$, we can now provide their graphical representation.

\begin{figure*}[t!]
    \centering
    \includegraphics[scale=0.72]{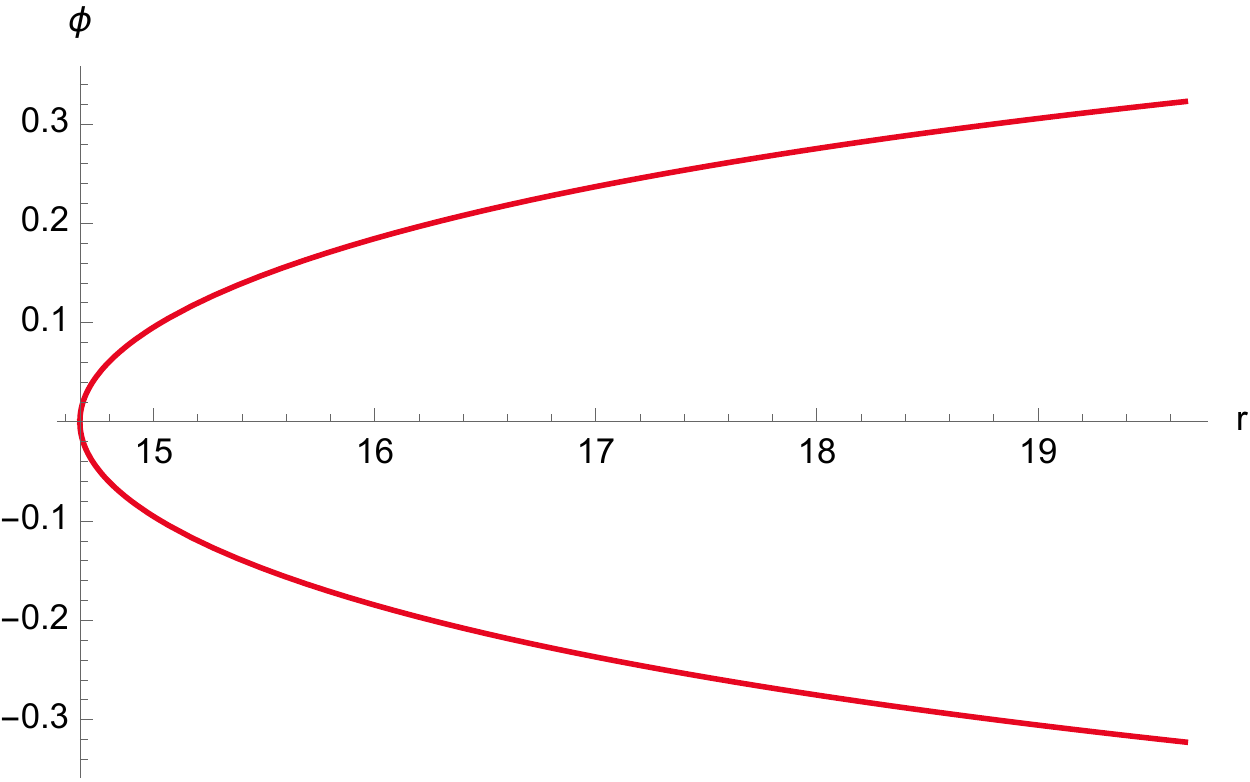}
    \caption{The function $\phi=\phi(r)$ for first-kind orbits having $C^2E^2<1$.    
The following constants have been chosen: $\phi_0=0$, $M=2$, $e=4.5$,  $\ell=11$, 
$\varepsilon= \pm 1$, and $C=1$.}
    \label{fig:E^2<1-first-kind-orbit}
\end{figure*}

\begin{figure*}[t!]
    \centering
    \includegraphics[scale=0.72]{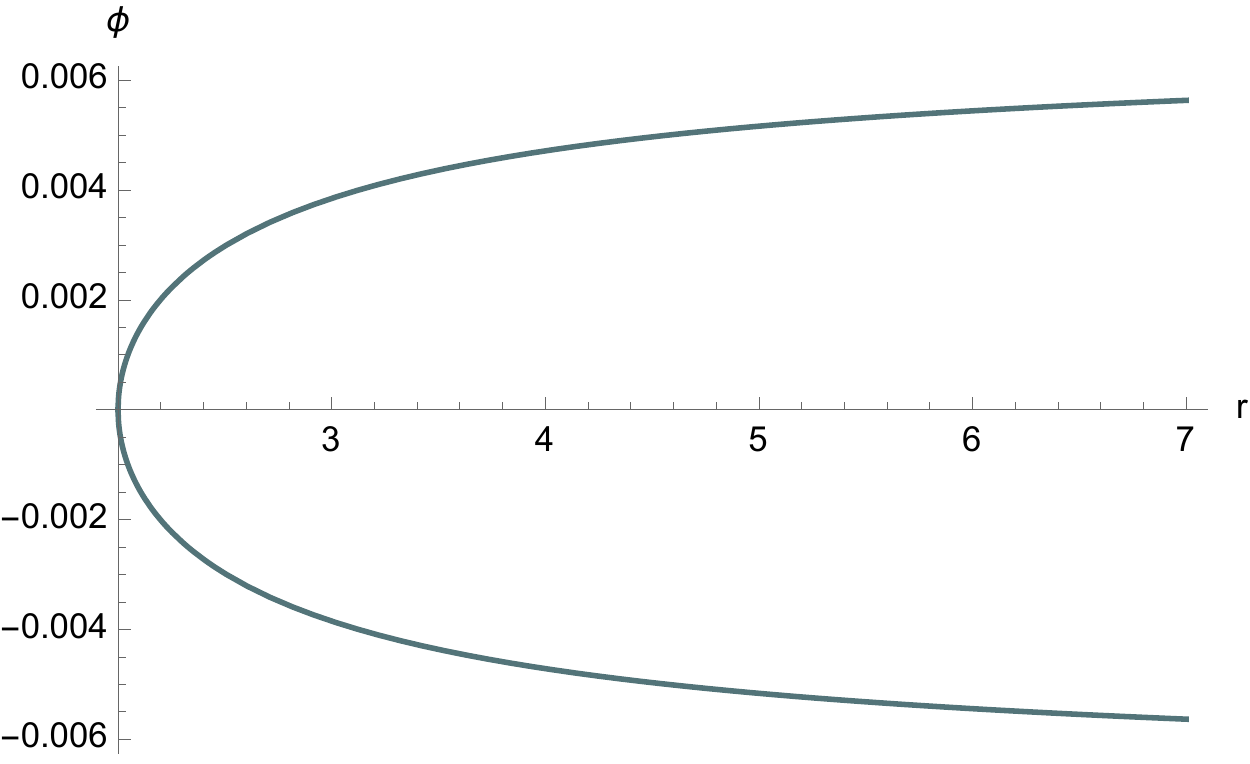}
    \caption{The function $\phi=\phi(r)$ for first-kind orbits having $C^2E^2<1$ in the limiting case (\ref{limiting-case}). The following 
constants  have been chosen: $\phi_0=0$, $M=1$, $e=1.5 \times 10^7$,  $\ell=10^5$, 
$\varepsilon= \pm 1$,   and $C=1$.}
    \label{fig:E^2<1-first-kind-orbit-limiting-case}
\end{figure*}

The  plot of the solution $\phi=\phi(r)$ for unbounded first-kind orbits
is displayed in Fig. \ref{fig:E^2<1-first-kind-orbit}, whereas the case of  the limiting regime (\ref{limiting-case}) is 
shown in Fig. \ref{fig:E^2<1-first-kind-orbit-limiting-case}.  It is clear that 
the resulting trajectory has the same behaviour as the orbit displayed in 
Fig. \ref{fig:E^2<1-first-kind-orbit}.

It should be noted that the limiting scenario (\ref{limiting-case}) is ruled by (cf. Eq. (\ref{u2-E^2<1}))
\begin{equation}
\mu \to 0. 
\label{limiting-1}
\end{equation}
By virtue of the constraint (\ref{mu<1/4}), the condition (\ref{limiting-1}) is admissible 
provided that (see Eq. (\ref{e>1}))
\begin{equation}
e \to + \infty,
\end{equation} 
whereas the definition (\ref{mu-definition}) of the parameter $\mu$ 
further demands (see Eq. (\ref{l>0}))
\begin{equation}
\ell \to + \infty.
\end{equation}
For the numerical evaluation of the inverse function $r=r(\phi)$, we refer
the reader to the method in Sec. III of Ref. \cite{Hackmann2008}. 
\begin{figure*}[t!]
    \centering
    \includegraphics[scale=0.72]{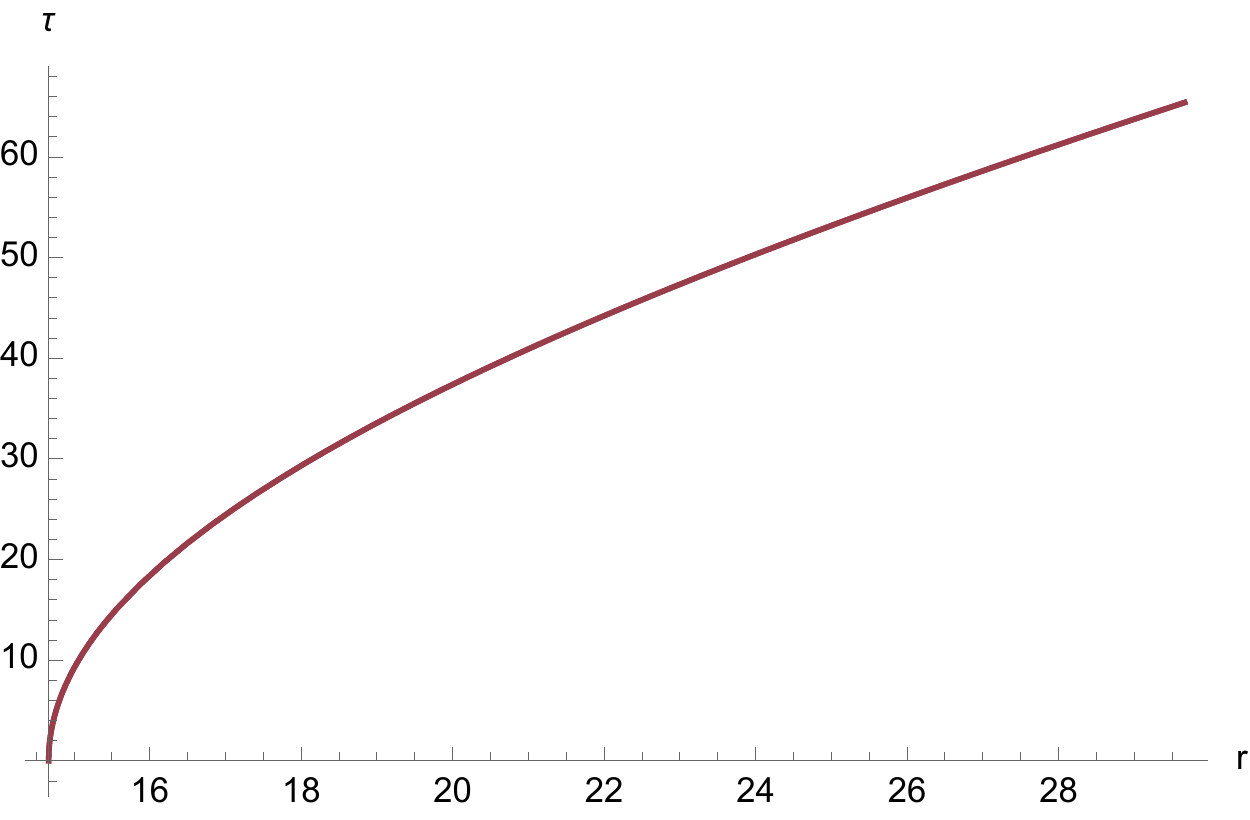}
    \caption{The function $\tau=\tau(r)$ for first-kind orbits having $C^2E^2<1$.    
The following constants have been chosen: $\tau_0=0$, $M=2$, $e=4.5$,  $\ell=11$, 
$\varepsilon= \pm 1$, and $C=1$.}
    \label{fig:E^2<1-first-kind-orbit-tau-of-r}
\end{figure*} 
\begin{figure*}[t!]
    \centering
    \includegraphics[scale=0.72]{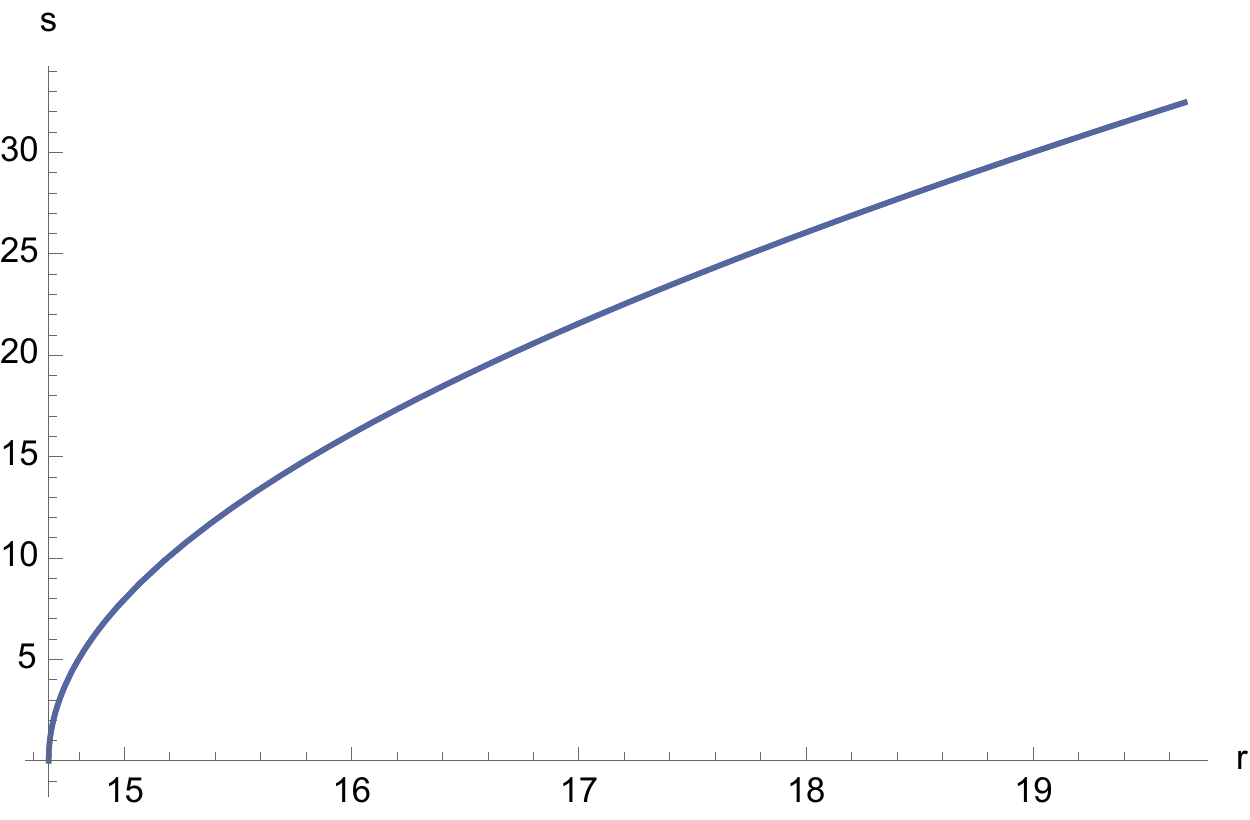}
    \caption{The function $s=s(r)$ for first-kind orbits having $C^2E^2<1$.    
The following constants have been chosen: $s_0=0$, $M=2$, $e=4.5$,  $\ell=11$, 
$\varepsilon= \pm 1$, and $C=1$.}
    \label{Fig-E-2-1-first-kind-orbit-s-of-r}
\end{figure*} 

The plots of the functions  $\tau=\tau(r)$ and $s=s(r)$ are given in Figs. \ref{fig:E^2<1-first-kind-orbit-tau-of-r} and \ref{Fig-E-2-1-first-kind-orbit-s-of-r}, respectively.

\subsection{Geodesics with $C^2E^2 \geq 1$} 

As pointed out before, as soon as $C^2E^2 >1$ the cubic \eqref{cubic-equation-F(u)=0} 
has only one positive root. We have  checked that this solution is always bigger than $1/2M$ (see also   Eq. \eqref{u1+u2+u3}). Therefore, in view of the constraint 
\eqref{u<1/2M},  no geodesic motion is allowed when $C^2E^2 >1$. In other words, 
no bounded orbit exists in Euclidean Schwarzschild geometry.

The condition (\ref{u<1/2M}) demands that the case $C^2E^2=1$ entails only the root 
$u=0$.  This means that when 
$C^2E^2=1$ the geodesic motion only allows $r=+\infty$. 

\section{Lack of circular orbits}\label{Sec:lack-circular-orbits}

The last interesting topic to be addressed concerns the investigation of the possible presence of circular orbits. This task is performed in this section, where we will consider $C=1$ for simplicity. 

\begin{figure*}[t!]
    \centering
    \includegraphics[scale=0.72]{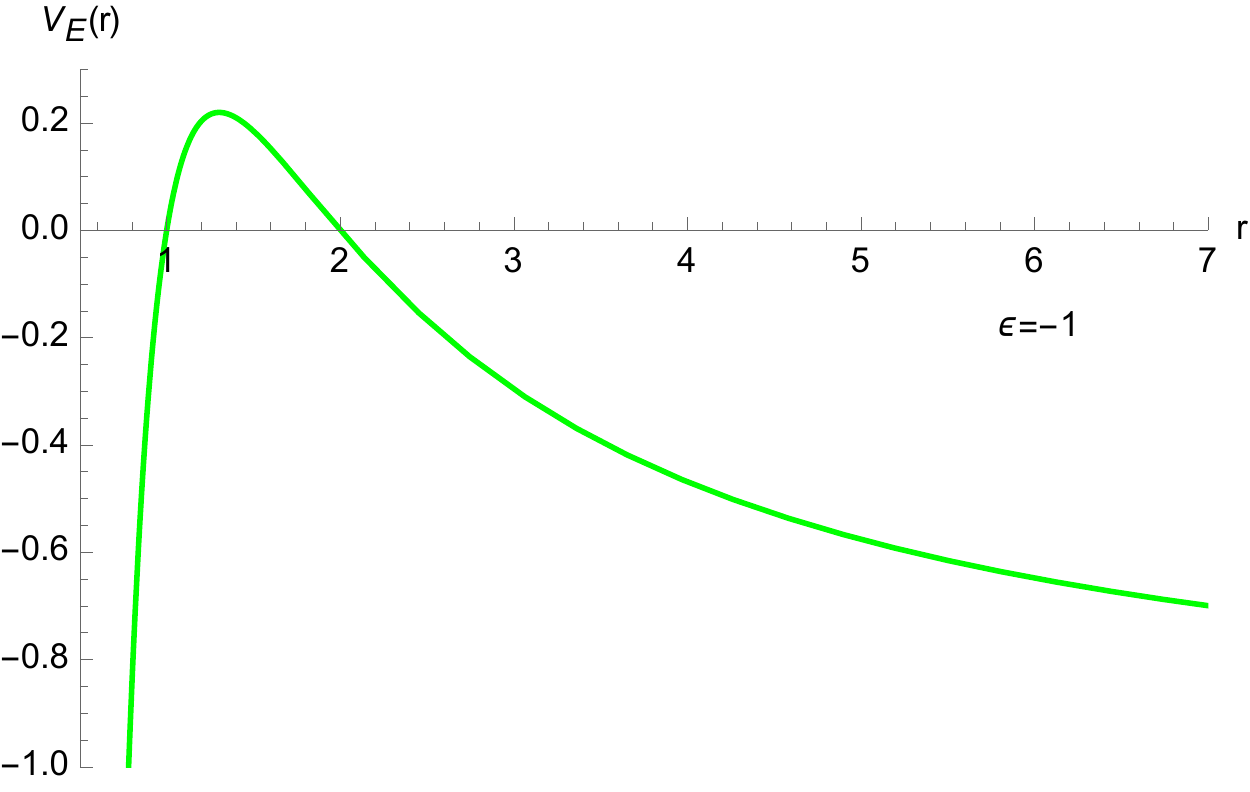}
    \caption{The potential energy function \eqref{euclidean-potential-energy} 
with $\varepsilon=-1$, $M=1$, and $L=1$. }
 \label{fig:potential-epsilon-minus1}
\end{figure*}

\begin{figure*}[t!]
    \centering
    \includegraphics[scale=0.72]{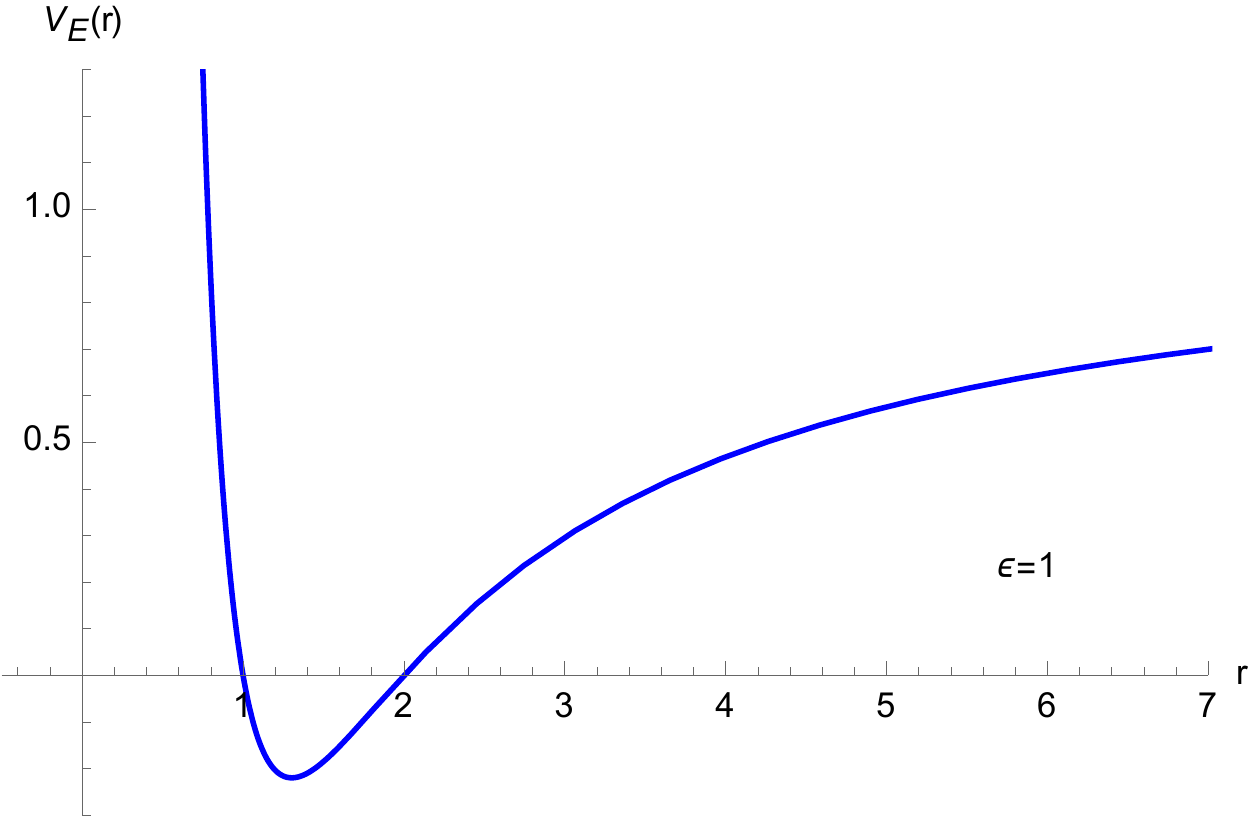}
    \caption{The potential energy function \eqref{euclidean-potential-energy} 
with $\varepsilon=1$, $M=1$, and $L=1$. }
 \label{fig:potential-epsilon-plus1}
\end{figure*}

By virtue of Eq. \eqref{Schwarzschild-geod-eqs-1-variable-r}, we can 
define an \qm{Euclidean potential energy} $V_E(r)$ as
\begin{align}\label{euclidean-potential-energy}
V_E(r)= \varepsilon \left(1-\dfrac{2M}{r}\right) \left(1-\dfrac{L^2}{r^2}\right),
\end{align}
where, as before, $\varepsilon=\pm 1$. It is known 
that \cite{Chandrasekhar1983} the minimum of the potential corresponds to 
a stable circular orbit, the maximum to an unstable one, whereas the point 
of inflection represents the innermost stable circular orbit. For the potential 
\eqref{euclidean-potential-energy}, we find that the first derivative 
\begin{align}
\dfrac{\dd V_E(r)}{\dd r}= \dfrac{2 \varepsilon}{r^4} \left[ M r^2 + L^2 (r-3M)\right],
\end{align}
vanishes at
\begin{align}
r_{1,2} &= \dfrac{-L^2  \mp \sqrt{L^4+12M^2 L^2}}{2M}.
\end{align}
Since $r_1<0$, we will only consider the solution 
\begin{align}
r_2 \equiv r^\star.
\end{align}
Then, from the study of the second derivative of $V_E(r)$, we obtain
\begin{align}
\left. \dfrac{\dd^2 V_E(r)}{\dd r^2}\right \vert_{r^\star}=\dfrac{32 \varepsilon 
M^4 L^2 \left(L^2+12M^2 -\sqrt{L^4+12M^2L^2}\right)}{\left(\sqrt{L^4+12M^2L^2}-L^2\right)^5},
\end{align}
which means that
\begin{align}
& r^\star \;  \mbox{ is a maximum of } V_E(r) \; \mbox{ if }   \varepsilon=-1,
\nn \\
& r^\star \;  \mbox{ is a minimum of } V_E(r) \; \mbox{ if }   \varepsilon=1,
\label{r-star-max-min}
\end{align}
as shown in Figs. \ref{fig:potential-epsilon-minus1} and \ref{fig:potential-epsilon-plus1}.
Furthermore, $r=r^\star$ cannot represent a point of inflection since the 
condition $\left. \tfrac{\dd^2 V_E(r)}{\dd r^2} \right \vert_{r^\star}=0$  
implies $12M^2 + L^2=0$, which in turn does not lead to any real-valued 
solution. Despite the result \eqref{r-star-max-min}, no circular orbit can 
exist in our model (even if $r^\star >2M$ when $\vert L \vert >2M$). In fact, 
bearing in mind Eq. \eqref{Schwarzschild-geod-eqs-1-variable-r}, we see that  
the requirement $\left(\tfrac{\dd r}{\dd s}\right)^2 >0$ entails, when $E^2>1$,
\begin{align}
\dfrac{V_E(r)}{\varepsilon} >1,
\end{align} 
but this lower bound is not fulfilled at $r=r^\star$. Furthermore, as a consequence of Eq. 
\eqref{Schwarzschild-geod-eqs-1-variable-r},  the condition 
$\left(\tfrac{\dd r}{\dd s}\right)^2 =0$ yields
\begin{align} \label{radial-geod-relation-V-E}
\dfrac{V_E(r)}{\varepsilon} =E^2,
\end{align} 
which, when evaluated at $r=r^\star$, leads  to complex-valued solutions for the energy $E$ (equivalently, 
these solutions do not satisfy  $E^2>1$ nor do they fulfill $0<E^2<1$)\footnote{The equation 
$\tfrac{V_E(r^\star)}{\varepsilon} =E^2$ leads also to the real-valued solution $E=0$ when $\vert L \vert =2M$. 
However, this solution cannot be accepted for two reasons: 
(i) it violates Eq. \eqref{E-definition}; 
(ii) it does not fulfill the constraint $r^\star >2M$.}. Since circular orbits 
are not present also if $E^2\leq 1$, 
this completes our proof that Euclidean 
Schwarzschild geometry does not envisage circular orbits. This differs from 
general relativity, where both stable and unstable circular trajectories are 
predicted, the innermost stable circular orbit occurring at 
$r=6M$  \cite{Chandrasekhar1983}.

We have been looking for circular geodesics that make a loop around the Euclidean time and correspond to constant values of $r$ and $\phi$. However, when $r$ and $\phi$ are constant, Eq. \eqref{Schwarzschild-geodetica-tau} is solved for $\tau(\lambda)$ by a linear function of the affine parameter, while Eq. \eqref{Schwarzschild-geodetica-r} shows that ${\rm d} \tau / {\rm d} \lambda =0$. Thus, the Euclidean time $\tau$ is found to be
constant, and the desired circular geodesic shrinks to the point
\begin{align}
(\tau= {\rm constant}, r={\rm constant}, \phi={\rm constant}).
\end{align}

\section{Conclusions}
\label{Sec:Conclusion}

In this paper we have evaluated in detail geodesic motion in Euclidean Schwarzschild geometry, 
limited to the real Riemannian section of the complexified Schwarzschild spacetime. 
Our explicit solution \eqref{(3.10)}--\eqref{(3.12)} in terms of incomplete elliptic integrals of first, second
and third kind has never appeared in the literature, to the best of our knowledge.

Our investigation has revealed new interesting features, which do not occur in the corresponding 
Lorentzian-signature framework. This means that the Euclidean and the Lorentzian Schwarzschild geometries 
are characterized by deep differences which cannot be merely reduced to the opposite signs occurring 
in the timelike component of their metric tensors. Indeed, we have shown that no elliptic-like orbits 
occur in the Euclidean Schwarzschild spacetime and, in general, bounded orbits are not allowed. 
Furthermore, unbounded orbits consist of first-kind trajectories only and are described by means 
of a parametrization which differs from the one adopted in general relativity (see Eq. \eqref{u1,u2,u3-E^2<1}). 

Recently, a new examination of the geodesic motion in Lorentzian Schwarzschild geometry has been proposed in the literature, where all kinds of nonradial causal geodesic orbits have been described via a single formula  making use  of  Weierstrass elliptic functions \cite{Cieslik2022}. On the other hand,  the Euclidean case studied in this paper exploits  incomplete elliptic integrals. Thus, an interesting issue to be addressed could consists in verifying whether the pattern of Ref. \cite{Cieslik2022} can be employed also in Euclidean settings.  

The lack of bounded orbits in Euclidean Schwarzschild geometry is a feature 
existing also at quantum level. Indeed, 
it has been shown in  Ref. \cite{S2016} that only the inclusion 
of a \qm{magnetic field} (i.e., a self-dual Abelian gauge field) yields bounded  
(elliptic) orbits (the same conclusions hold also for Taub-NUT and Taub-Bolt 
spaces, see Refs. \cite{Schroers2016a,Schroers2021}). Moreover, in this framework 
(and in particular in the context of the recently proposed geometric models of matter 
\cite{Atiyah2012}) the Euclidean Schwarzschild space emerges as a natural   
geometric candidate for the neutron \cite{S2015} (whereas the Euclidean 
Taub-NUT space can represent the electron \cite{Atiyah2012}).  

The investigation of singularities in Euclidean Schwarzschild geometry is a 
physical motivation supporting our paper. In fact, it is known \cite{Hawking-Ellis} 
that in general relativity timelike and null geodesic incompleteness is the 
criterion used to define the occurrence of space-time singularities.  
On the other hand, in the case of Euclidean Schwarzschild geometry, the 
absence of the singularity at $r=0$ is demonstrated via a \qm{shortcut} by 
considering the real section of the complexified Schwarzschild spacetime in  
Kruskal-Szekeres coordinates \cite{GH1977}. Our analysis can be 
thus exploited to show that the geodesics of (the real section of) the 
Euclidean Schwarzschild spacetime are indeed complete and hence no singularity 
can emerge. 

Last,  this work can represent a starting point for a systematic study of geodesic motion in Euclidean gravity. 
Thus, the first step carried out in this paper can be followed by an analysis involving the whole set 
of gravitational instantons in general. This might entail the discovery of new results both in Riemannian 
geometry and Euclidean quantum gravity. 

\section*{Acknowledgements}
This work is supported by the Austrian Science Fund (FWF) grant P32086.

\begin{appendix}

\section{General formulae for roots of the cubic $\mathcal{F}(u)=0$}
\label{Appendix:general-formulae-roots}

The three roots of the cubic  \eqref{cubic-equation-F(u)=0} can be obtained 
by means of a \emph{numerical} evaluation of the following quantities 
(recall that $m \equiv M/L$):
\begin{align}
u_1 &=\dfrac{1}{6M} +\dfrac{\left(1+ {\rm i} \sqrt{3}\right)\left(1+12m^2\right)}{\left(6M \sqrt[\leftroot{0}\uproot{2}3]{4} 
\right)\mathscr{B}} +\dfrac{\left(-1+{\rm i}\sqrt{3}\right)\mathscr{B}}{12M \sqrt[\leftroot{0}\uproot{2}3]{2}}, 
\\
u_2 &=\dfrac{1}{6M} -\dfrac{\left(1- {\rm i} \sqrt{3}\right)\left(1+12m^2\right)}{\left(6M \sqrt[\leftroot{0}\uproot{2}3]{4} 
\right)\mathscr{B}} -\dfrac{\left(1+{\rm i}\sqrt{3}\right)\mathscr{B}}{12M \sqrt[\leftroot{0}\uproot{2}3]{2}}.
\\
u_3 &= \dfrac{1}{6M} +\dfrac{\left(1+12m^2\right)}{\left(3M \sqrt[\leftroot{0}\uproot{2}3]{4} 
\right)\mathscr{B}} +\dfrac{\mathscr{B}}{6M \sqrt[\leftroot{0}\uproot{2}3]{2}}, 
\end{align} 
where 
\begin{align}
\mathscr{B}  \equiv \sqrt[\leftroot{0}\uproot{2}3]{\left(2-72m^2+108C^2E^2 m^2 \right) 
+\sqrt{\left(2-72m^2+108C^2E^2m^2\right)^2-4 \left(1+12m^2\right)^3}}.
\end{align}

The above formulae are valid for any real-valued $C$ and $E$. 

\section{More details about the roots of the cubic $\mathcal{F}(u)=0$ under the hypothesis $C^2E^2<1$}
\label{Sec:roots-cubic-E^2<1}

In Sec. \ref{Sec:Sol-elliptic.intregrals}, we have seen that the form  (\ref{u1,u2,u3-E^2<1}) of 
the roots of the cubic (\ref{cubic-equation-F(u)=0})   accounts correctly for the 
geodesic motion under the hypothesis  $C^2E^2<1$. In this Appendix we will show that, had 
we chosen the same parametrization as in general relativity \cite{Chandrasekhar1983}, 
we would have obtained some inconsistencies. Let 
\begin{subequations}
\label{u1,u2,u3-E^2<1-appendix}
\begin{align}
u^\prime_1 &= -\dfrac{1}{\ell} \left( e-1\right), 
\label{u1-E^2<1-appendix}
\\
u^\prime_2 &= \dfrac{1}{\ell} \left( e+1\right), 
\label{u2-E^2<1-appendix}
\\
u^\prime_3 &= \dfrac{1}{2M} -\dfrac{2}{\ell},
\label{u3-E^2<1-appendix}
\end{align}
\end{subequations}
denote a new (hypothetical) set of  roots of the cubic (\ref{cubic-equation-F(u)=0}) in the case $C^2E^2<1$. 
Equation (\ref{u1,u2,u3-E^2<1-appendix}) is identical to the choice adopted in the 
context of Einstein's theory \cite{Chandrasekhar1983} and furthermore it is clear that 
(cf. Eq. (\ref{u1,u2,u3-E^2<1})) $u^\prime_1=u_1$, $u^\prime_2=u_3$, and $u^\prime_3=u_2$. 
In addition, by means of the parametrization (\ref{u1,u2,u3-E^2<1-appendix}), the 
constraint (\ref{u1,u,2u3-E^2<1-inequalities}) is not enforced since
\begin{equation} \label{u1,u,2u3-E^2<1-inequalities-appendix}
u^\prime_1 <0 <u^\prime_2 < u^\prime_3 <\dfrac{1}{2M},
\end{equation}
provided that
\begin{subequations}
\label{constraints_1-appendix}
\begin{align}
\ell & >0, 
 \\
e & >1,
\label{e>1-appendix}
 \\
\mu &<\dfrac{1}{4},
\label{mu<1/4-appendix}
 \\
1-6\mu -2\mu e &> 0,
\label{1-6mu-2mue>0-appendix}
\end{align}
\end{subequations}
where $\mu$ has been defined in Eq. (\ref{mu-definition}). Condition (\ref{mu<1/4-appendix}) is 
guaranteed by (\ref{e>1-appendix}) and (\ref{1-6mu-2mue>0-appendix}).  Vi\`{e}te's formulae  
(\ref{u1u2 + u1u3 + u2u3}) and (\ref{u1u2u3}) lead to the same relations as in Eq. 
(\ref{relations-1/L^2-E^2<1}), i.e., 
\begin{subequations}
\label{relations-1/L^2-E^2<1-appendix}
\begin{align}
\dfrac{1}{L^2} &= \dfrac{\mu \left(3+e^2\right)-1}{M \ell},
\label{inequality-1/L^2-appendix}
\\
\dfrac{\left(1-C^2E^2\right)}{L^2} &= \dfrac{\left(e^2-1\right)\left(1-4 \mu \right)}{\ell^2}.
\end{align}
\end{subequations}
Equations (\ref{1-6mu-2mue>0-appendix}) and (\ref{inequality-1/L^2-appendix})  entail
\begin{align}
e>3, \qquad \dfrac{1}{\left(3+e^2\right)}< \mu < \dfrac{1}{\left(6+2e\right)},
\end{align}
and hence the set of constraints (\ref{constraints_1-appendix}) should be slightly  modified according to
\begin{subequations}
\label{constraints_2-appendix}
\begin{align}
\ell & >0, 
 \\
e & >3,
\label{e>3-appendix}
 \\
\mu &<\dfrac{1}{4},
\label{mu<1/4_2-appendix}
 \\
\dfrac{1}{\left(3+e^2\right)} &< \mu < \dfrac{1}{\left(6+2e\right)},
\label{mu-range-appendix}
\end{align}
\end{subequations}
where (\ref{mu<1/4_2-appendix}) is fulfilled on account of (\ref{e>3-appendix}) and 
(\ref{mu-range-appendix}). Note that, unlike the analysis of  Sec. \ref{Sec:Sol-elliptic.intregrals}, 
the parameter $e$ is such that $e \notin (1,3)$ (cf. Eqs. (\ref{e>1}) and (\ref{e>3-appendix})). 
At this stage, from the requirement 
\begin{equation}
E^2>0,
\end{equation}
we obtain from Eq. (\ref{relations-1/L^2-E^2<1-appendix})
\begin{equation}
\mu > \dfrac{1}{2\left(e+1\right)},
\end{equation}
which disagrees with Eq. (\ref{mu-range-appendix}). Thus, as we have seen in Sec. 
\ref{Sec:Sol-elliptic.intregrals}, the parametrization (\ref{u1,u2,u3-E^2<1}) should be employed 
in place of (\ref{u1,u2,u3-E^2<1-appendix}).

\section{Incomplete elliptic integrals}\label{Appendix:Incompl-integr}

According to the standard notation in Ref. \cite{Byrd}, the incomplete
elliptic integrals of the first, second, and third kind can be defined as
\begin{align}
F(\varphi,k^{2}) &=\int_{0}^{\varphi}
{\dd \theta \over \sqrt{1-k^{2}\sin^{2}\theta}},
\label{(A1)}
\\
E(\varphi,k^{2}) &=\int_{0}^{\varphi}
\sqrt{1-k^{2}\sin^{2}\theta} \; \dd \theta,
\label{(A2)}
\\
\pi(\varphi,\alpha^{2},k^{2})&=\int_{0}^{\varphi}
{\dd \theta \over (1-\alpha^{2}\sin^{2}\theta)
\sqrt{1-k^{2}\sin^{2}\theta}},
\label{(A3)}
\end{align}
 respectively, while the Jacobi elliptic functions read as
\begin{equation}
{\rm sn}(u)=\sin \varphi, \quad
{\rm cn}(u)=\cos \varphi, \quad
{\rm dn}(u)=\sqrt{1-k^{2}\sin^{2}\varphi}.
\label{(A4)}
\end{equation}

\end{appendix}

\end{document}